\documentclass[prc,twocolumn,showpacs,preprintnumbers,amsmath,amssymb]{revtex4}
\usepackage{graphicx}
\usepackage{dcolumn}
\usepackage{bm}
\def\be{\begin{equation}}
\def\ee{\end{equation}}
\def\ba{\begin{array}}
\def\ea{\end{array}}
\def\bea{\begin{eqnarray}}
\def\eea{\end{eqnarray}}
\begin{document}
\title{Dynamical cluster-decay model for hot and rotating light-mass
nuclear systems, applied to low-energy
$^{32}$S+$^{24}$Mg$\rightarrow ^{56}$Ni$^*$ reaction}

\author{Raj K. Gupta$^1$, M. Balasubramaniam$^{1,2}$, Rajesh Kumar$^1$, Dalip Singh$^1$, C. Beck$^3$, and Walter Greiner$^4$}

\affiliation{$^1$Department of Physics, Panjab University,
Chandigarh-160014, India.\\
$^2$Department of Physics, Manonmaniam Sundaranar University,
Tirunelveli-627012, India.\\
$^3$Institut de Recherches Subatomiques, UMR7500, IN2P3/CNRS - Universit\'e Louis Pasteur, B.P.~28, F-67037 Strasbourg Cedex 2, France.\\
$^4$Institut f\"ur Theoretische Physik, J-W Goethe Universit\"at,
D-60054 Frankfurt/Main, Germany.}

\date{\today}
\begin{abstract}
The dynamical cluster-decay model (DCM) is developed further for the decay of hot and rotating compound nuclei (CN) formed in light 
heavy-ion reactions. The model is worked out in terms of only one
parameter, namely the neck-length parameter, which is related to
the total kinetic energy TKE(T) or effective Q-value Q$_{eff}(T)$ at
temperature T of the hot CN, defined in terms of the CN binding
energy and ground-state binding energies of the emitted fragments. The emission of both the light-particles (LP), with A$\le $4, Z$\le $2, as well as the complex intermediate mass fragments (IMF), with 
4$<$A$<$20, Z$>$2, is considered as the dynamical collective
mass motion of preformed clusters through the barrier. Within the same
dynamical model treatment, the LPs are shown to have different
characteristics as compared to the IMFs. The systematic variation of the LP emission cross section $\sigma_{LP}$, and IMF emission 
cross section $\sigma_{IMF}$, calculated on the present DCM match exactly the statistical fission model predictions. It is for the first time that a non-statistical dynamical description is developed for the emission of light-particles from the hot and rotating CN. The model is applied to the decay of $^{56}Ni^*$ formed in the $^{32}$S+$^{24}$Mg reaction at two incident energies $E_{c.m.}$=51.6 and 60.5 MeV. Both the IMFs and average $\overline{TKE}$ spectra are found to compare resonably nicely with the experimental data, favoring asymmetric mass distributions. The LPs emission cross section is shown to depend strongly on the type of emitted particles and their multiplicities.
\end{abstract}


\maketitle

\section{INTRODUCTION}

Light compound nuclei (CN) formed in low energy (E/A$<$15 MeV/nucleon) heavy-ion reactions are highly excited and carry large angular momenta. The compound systems so formed decay by emitting multiple light particles (n, p, $\alpha$) and $\gamma$-rays, which in the statistical Hauser-Feshbach (HF) analysis is understood as an emission process from the equilibrated CN, resulting in CN fusion cross sections. Apparently, the decay process must depend on temperature and angular momentum dependent potential barriers ~\cite{sanders99}. For light compound systems with $A_{CN}\ge 40$, the above noted light-particles (LP) emission is always accompanied by intermediate mass fragments, the IMFs of Z$>$2 and 4$<$A$<$20, also called "complex fragments" or "clusters", whose contribution, though small of the order of 5 to 10\%, is to be included in the CN fusion cross sections. Then, the temperature- and spin-dependent potentials must also be mass-asymmetry dependent. In other words, the structure effects of the compound system also become important.

In order to understand the IMF production, not only the HF analysis is extended to include the fragments heavier than $\alpha$-particle in the BUSCO code \cite{campo91} or in the Extended Hauser-Feshbach 
scission-point model \cite{matsuse97}, but also other statistical fission model descriptions~\cite{sanders99} have been used that are based on either the scission-point or saddle-point configuration, in the GEMINI code \cite{charity88} or 
the saddle-point "transition-state" model (TSM)
\cite{moretto75,sanders99,sanders89,sanders91,nouicer99}, respectively. The LP emission in these fission models is still treated within the statistical HF method. It is interesting to mention that light-nucleus emission can also be qualitatively well described in the framework of a generalized rotating liquid-drop model, proposed recently by Royer and collaborators \cite{royer02,royer03a,royer03b}. Since the measured angular distributions and energy spectra of emitted complex fragments are consistent with fission-like decays of the respective compound
system, the fusion-fission process is now well established in light dinuclear systems~\cite{sanders99}. The statistical fission models, for
which the fission-decay of a CN is determined by the phase space (level density) available at the saddle-point~\cite{charity88,sanders91}
(or scission-point~\cite{matsuse97}) configuration, may however lack in
terms of not including more explicitly the structure effects of the compound system. Large structure effects have been shown to be important in the $^{56}$Ni compound system through a strong resonance behaviour~\cite{beck94,beck95} of the excitation functions of
large-angle $^{28}$Si+$^{28}$Si elastic and inelastic scattering
yields~\cite{nouicer99,beck01}. Although neither similar resonant effects nor orbiting processes~\cite{shapira88} have been evidenced in the $^{32}$S+$^{24}$Mg reaction~\cite{sanders87,sanders88,sanders90} studied here, a fully dynamical theory for more complete description
of the emission of both the LPs and IMFs in the framework of the
statistical model of the decay of such a hot and rotating nuclear 
$^{56}$Ni system remains highly desirable.

A recent attempt towards the dynamical treatment of the decay of a
hot and rotating nucleus is the work of Gupta and collaborators
\cite{sharma00,gupta02,gupta03,bala03,gupta03b,gupta04}, where a dynamical collective clusterization process is proposed as a possible alternative of the fission process. Both the LPs and IMFs are considered as the dynamical mass motion of preformed fragments or clusters through the barrier. Note that, in terms of the barrier picture, a cluster-decay process is in fact a fission process with structure effects of the CN also included via the preformation of the fragments, but without any phase space arguments (i.e. with no level density calculations). Alternatively, the dynamical fission process has been considered by some authors \cite{maruhn74,gupta75} simply as a continuous deformation of the CN
\cite{gupta84,saroha85,malik86,puri92,gupta97a}. The above noted dynamical cluster-decay model (DCM) of Gupta {\it et al.}
\cite{sharma00,gupta02,gupta03,bala03} is so-far used to calculate the
decay constant and total kinetic energies (TKEs) of IMFs alone, and that too in a re-normalization procedure only. In other words, only the angular momentum $\ell$=0 (s-wave) solution and TKE for fixed 
$\ell$-value are considered. In this paper, we develop the DCM further for the calculation of actual (summed up, "total" $\ell$) cross sections and the average total kinetic energies $\overline{TKE}s$ for both the emitted light particles and complex IMFs. Interesting enough, the light particles, though treated within the same dynamical collective clusterisation process, are found to possess different characteristic properties. Brief reports of the present work have recently been presented elsewhere~\cite{gupta03b,gupta04}.

The data for IMFs chosen here, to apply the DCM, are those of the
$^{32}$S+$^{24}$Mg$\rightarrow ^{56}$Ni$^*$ reaction, where the mass spectra for A = 12 to 28 fragments, and the average total kinetic energy ($\overline{TKE}$) for only the most favored (enhanced yields)
$\alpha$-nucleus fragments, are measured at two incident energies
$E_{lab}$ = 121.1 and 141.8 MeV, or equivalently at $E_{c.m.}$ = 51.6
and 60.5 MeV, respectively \cite{sanders89,sanders87} (only even-even fragments are observed for $E_{c.m.}$ = 51.6 MeV). In this experiment, by detecting both fragments in coincidence, it was possible to deduce the primary mass distribution for the decay process i.e. the mass distribution before the occurrence of the secondary light-particles emission from the fragments. These primary, pre-evaporation mass distributions show that the mass-asymmetric channels are favored over the symmetric ones, with $\alpha$-nucleus, A=4n, fragments having enhanced yields. The IMF emission cross section, estimated in one of these experiments \cite{sanders87}, is ${59\pm 12}$ mb. The CN fusion cross section data due to multiple LP emission at these energies are also deduced later from the same experiment \cite{sanders89}, and as quoted in \cite{sanders91}, are $1080\pm 130$ and $1050\pm 100$ mb, respectively, at $E_{c.m.}$ = 51.6 and 60.5 MeV, which fit the other available earlier measurements at similar and other energies for this system \cite{sanders87,hinnefeld87,gutbrod73}. The total fusion cross section is then the sum of this cross section due to the LP emission and the fission-like IMF emission cross section.

The paper is organised as follows. Section II gives briefly a description of the DCM for hot and rotating light-mass nuclei.
The model is presented for the emission of both the LPs and IMFs. The
calculations are presented in section III. The discussion of our results and a summuary constitute section IV.


\section{THE DYNAMICAL CLUSTER-DECAY MODEL FOR HOT AND ROTATING COMPOUND SYSTEMS}

The DCM for hot and rotating nuclei is a reformulation of the preformed cluster model (PCM) of Gupta and collaborators for ground-state decays in cluster radioactivity (CR) and related phenomena
\cite{gupta88,malik89,gupta91,kumar94,gupta94,gupta99a}. Thus, like PCM, DCM is also based on the dynamical (or quantum mechanical)
fragmentation theory of cold fusion phenomenon in heavy ion reactions and fission dynamics \cite{maruhn74,gupta75,gupta77,maruhn80,gupta99b},
including the prediction of CR \cite{gupta94,sandu80,bonetti99}. This theory is worked out in terms of:\\
(i) the collective coordinate of mass (and charge) asymmetry (H and L
stand, respectively, for heavy and light fragments)
$$\eta={{(A_H-A_L)}/{(A_H+A_L)}}$$
$$(\hbox{and}\quad\eta _Z={{(Z_H-Z_L)}/{(Z_H+Z_L)}}),$$
and (ii) relative separation R,\\
which in DCM characterizes, respectively, \\
(i) the nucleon-division (or -exchange) between outgoing fragments,
and \\
(ii) the transfer of kinetic energy of incident channel ($E_{c.m.}$) to
internal excitation (total excitation or total kinetic energy, TXE or TKE) of the outgoing channel, since the fixed $R=R_a$ (defined later), at which the process is calculated, depends on temperature T as well as on $\eta$, i.e. $R(T,\eta)$. This energy transfer process follows the relation (see Fig. 1)
\be
E_{CN}^*=E_{c.m.}+Q_{in}=\mid Q_{out}(T)\mid+TKE(T)+TXE(T).
\label{eq:1}
\ee
Note that for the $^{32}$S+$^{24}$Mg$\rightarrow ^{56}$Ni$^*$ reaction,
$Q_{in}$ is positive (=16.68 MeV) and hence adds to the entrance channel kinetic energy $E_{c.m.}$ of the two incoming nuclei in their ground states, and $Q_{out}(T)$ is negative and different for different exit channels at a fixed temperature T. The CN excitation energy 
$E_{CN}^*$ and its temperature $T$ (in MeV) are related as
\be
E_{CN}^*=\left ({A/9}\right ){T}^2-T.
\label{eq:2}
\ee

Using the decoupled approximation to R- and $\eta$-motions, the DCM defines the decay cross section, in terms of partial waves, as
\cite{gupta03,bala03}
\be
\sigma=\sum_{\ell=0}^{\ell_{max}}\sigma_{\ell}={\pi \over k^2}
\sum_{\ell=0}^{\ell_{max}}(2\ell+1)P_0P;
\qquad
k=\sqrt{2\mu E_{c.m.}\over {\hbar^2}}
\label{eq:3}
\ee
where, $P_0$, the preformation probability, refers to $\eta$-motion and
P, the penetrability, to R-motion. Apparently, for $\ell$=0 (s-wave)
$\sigma_{0}={\pi \over k^2}P_0P$, which is an equivalent of decay constant $\lambda=\nu_0P_0P$ (or decay half-life 
T$_{1/2}={ln 2}/\lambda$) with $\nu_0$ as the barrier assault frequency. In other words, $\sigma_{0}$ and $\lambda$ differ through a constant only. Thus, like in PCM, here the complex fragments (both LPs
and IMFs) are treated as the dynamical collective mass motion of
{\it preformed clusters or fragments} through the barrier. The structure information of the CN enters the model via the preformation
probabilities $P_0$ (also known as the spectroscopic factors) of the fragments.

The $P_0$ is given by the solution of stationary Schr\"odinger equation
in $\eta$, at a fixed R,
\be
\{ -{{\hbar^2}\over {2\sqrt B_{\eta \eta}}}{\partial \over {\partial
\eta}}{1\over {\sqrt B_{\eta \eta}}}{\partial\over {\partial \eta
}}+V(R,\eta ,T)\} \psi ^{\nu}(\eta ) = E^{\nu} \psi ^{\nu}(\eta ),
\label{eq:4}
\ee
with $\nu$=0,1,2,3... referring to ground-state ($\nu =0$) and
excited-states solutions. Then, the probability
\be
P_0(A_i)=\mid \psi (\eta (A_i))\mid ^2 {\sqrt {B_{\eta \eta}}} {2\over A},
\label{eq:5}
\ee
where $i$= 1 or 2, the heavy (H) and light (L) fragment, respectively,
and for a Boltzmann-like function
\be
\mid \psi \mid ^2 = \sum _{\nu =0}^{\infty}\mid\psi ^{\nu}\mid ^2
exp (-E^{\nu}/T).
\label{eq:6}
\ee

The constant $R=R_a=C_t=C_1+C_2$ in (\ref{eq:4}), fixed empirically
as the first turning point of the penetration path (like in Fig. 1) for
the ground state ({\it g.s}, T=0) decay, since this value of R (instead
of CN radius $R_0$) assimilates the effects of the deformations of two fragments and neck formation between them \cite{kumar97}. Thus, for the deformation effects included, we have for {\it g.s.} decay
\be
R_a(T=0)=C_t\quad {\hbox {or, in general,}}\quad =C_t+\sum _{i=1}^{2}\delta R(\beta_i),
\label{eq:7}
\ee
an equivalent of lowering of the barrier for deformed fragments
\cite{kumar97}. This is important here because the expected shapes of
some of the observed fragments in the exit channel of the reactions studied here could be deformed \cite{beck01,sanders94}. Thus, the deformation effects of the fragments (and the neck formation between them), included here within the extended model of Gupta and collaborators \cite{kumar97,khosla90,gupta97}, are treated via a neck-length parameter $\sum \delta R(\beta_i)$ at the scission configuration
which simulates the two centre nuclear shape parameterization, used
for both the light \cite{khosla90,gupta97} and heavy \cite{kumar97}
nuclear systems. This method of introducing a neck-length parameter
$\sum \delta R$ is similar to that used in both the scission-point
\cite{matsuse97} and saddle-point \cite{sanders89,sanders91} statistical fission models. The alternate, i.e., the calculation of fragmentation potential $V(\eta)$ and scattering potential V(R) for deformed nuclei, though shown to be difficult \cite{kumar97}, is being worked out \cite{gupta03a,gupta04x}. Also, we use here the S\"usmann central radii $C_i=R_i-{b^2/{R_i}}$ (in fm), where 
$R_i=1.28A_i^{1/3}-0.76+0.8A_i^{-1/3}$ fm and surface thickness
parameter $b=$0.99 fm. Note that the $C_t$ are different for different
$\eta$-values and hence $R_a(T=0)$ also depends on $\eta$.

For the decay of a hot CN, we use the postulate
\cite{gupta02,gupta03,bala03} for the first turning point
\be
R_a(T)=R_a(\eta , T=0)+\Delta R(\eta , T),
\label{eq:8}
\ee
depending on the total kinetic energy TKE(T). The corresponding potential $V(R_a)$ acts like an effective Q-value, $Q_{eff}$, for the decay of the hot CN at temperature T, to two exit-channel fragments
observed in {\it g.s.} T=0, defined by
\bea
Q_{eff}(T)&=&B(T)-[B_L(T=0)+B_H(T=0)]\nonumber\\
          &=&TKE(T)=V(R_a(T)).
\label{eq:9}
\eea
Here B's are the respective binding energies. Thus, $\Delta R$ in
Eq.(\ref{eq:8}) gives the change in TKE(T) {\it w.r.t.} TKE(T=0), which is taken to depend on $\eta$. As a first approximation to Eq.(\ref{eq:9}), in our earlier calculations \cite{gupta02,gupta03,bala03}, we used a constant average $\overline{\Delta R}$, which also takes care of the additional $\sum\delta R(\beta_i)$ effects of the deformations of fragments and neck formation between them,
\be
R_a(T)=C_t(\eta,T)+\overline{\Delta R}(T).
\label{eq:10}
\ee
Note that here $C_t$ is also taken to depend on temperature, to be
defined later. In the following calculations, we shall first use a
constant $\overline{\Delta R}(T)$ and then show the effect of using the
actual $\eta$-dependent $\Delta R(\eta , T)$, calculated from
Q$_{eff}$ or $V(R_a,T)$ defined by Eq.(\ref{eq:9}).

The above defined decay of a hot CN to two cold (T=0) fragments, via Eq.(\ref{eq:9}), could apparently be achieved only by emitting some light particle(s), like n, p, $\alpha$, or $\gamma$-rays of energy
\bea
E_x&=&B(T)-B(0)=Q_{eff}(T)-Q_{out}(T=0)\nonumber\\
   &=&TKE(T)-TKE(T=0)
\label{eq:11}
\eea
which is zero for the g.s. decay, like for exotic CR. Note that the
second equality in Eq. (\ref{eq:11}) is not defined for a negative
$Q_{out}(T=0)$ system since the negative TKE(T=0) has no meaning.
Apparently, Eq. (\ref{eq:11}) {\it w.r.t} (\ref{eq:9}) suggests that the emission of light-particles starts early in the decay process. The exit channel fragments in (\ref{eq:9}) are then obtained in the ground-state with TKE(T=0), as can be seen by calculating $E_{CN}^*-E_x$:
\be
E_{CN}^*-E_x=\mid Q_{out}(T)\mid+TKE(T=0)+TXE(T).
\label{eq:11a}
\ee
The excitation energy TXE(T) is used in, not treated here, the secondary emission of light particles from the fragments which are otherwise in their ground states with TKE(T=0) in the radial motion. Instead, we compare present calculations with the primary pre-secondary-evaporation fragment emission data~\cite{sanders89,sanders87}. Thus, by defining $Q_{eff}(T)$ as in (\ref{eq:9}), in this model we treat the LP emission at par with the IMF emission. In other words, a non-statistical dynamical treatment is attempted here for the first time for not only the emission of IMFs but also of multiple LPs, understood so-far only as the statistically evaporated particles in a CN emission. It may be reminded here that the statistical model (CN emission) interpretation of IMFs is not as good as it is for the LP production\cite{campo91,matsuse97,charity88,sanders89,sanders91,beck01}.

The collective fragmentation potential $V(R,\eta,T)$ in Eq.(\ref{eq:4}) is calculated according to the Strutinsky method by using the 
T-dependent liquid drop model energy $V_{LDM}$ of 
Ref. \cite{davidson94}, with its constants at T=0 re-fitted
\cite{gupta03,bala03} to give the recent experimental binding energies
\cite{audi95}, and the "empirical" shell corrections $\delta U$ of 
Ref. \cite{myers66} (In the Appendix of \cite{gupta03} and Eq. (8) of \cite{bala03}, $a_a$=0.5, instead of unity). Then, including  the 
T-dependence also in Coulomb, nuclear proximity, and $\ell$-dependent
potential in complete sticking limit of moment of inertia, we get
\bea
V(R,\eta ,T)&=&\sum_{i=1}^{2}{\Bigl [ V_{LDM}(A_i,Z_i,T)\Bigr ]}\nonumber\\
&+&\sum_{i=1}^{2}{\Bigl [ \delta U_i\Bigr ]}exp(-\frac{{T}^2}{{T_0}^2})
\nonumber\\
&+&E_c(T)+V_P(T)+V_{\ell}(T).
\label{eq:12}
\eea
The shell corrections $\delta U(T)$ in (\ref{eq:12}) are taken to go to
zero exponentially with T, with $T_0=1.5$ MeV \cite{jensen73}. The other T-dependent terms $E_c$, $V_P$ and $V_{\ell}$ are defined as follows:
\be
E_c(T)=Z_HZ_Le^2/R(T),
\label{eq:13}
\ee
where the charges $Z_i$ are fixed by minimizing the potential
$V(R,\eta ,T)$ in the charge asymmetry coordinate $\eta _Z$ (for fixed R and $\eta$ coordinates). Note, here the T-dependence of $E_c$ is the same as in Ref. \cite{davidson94}. Next,
\be
V_P(T)=4\pi\bar{R}(T)\gamma b(T)\Phi (s(T)),
\label{eq:14}
\ee
with $\bar{R}(T)$ and $\Phi (s(T))$, respectively, as the inverse of the root mean square radius of the Gaussian curvature and the universal
function, independent of the geometry of the system, given by
\cite{blocki77}
\be
\bar{R}(T)=\frac{C_1(T)C_2(T)}{C_t(T)},
\label{eq:16}
\ee
\be
\Phi (s(T))=\left \{
\ba{ll}
-{1 \over 2}(s-2.54)^2-0.0852(s-2.54)^3 ~s\le 1.2511 \\
-3.437exp(-{s \over 0.75}) ~s\ge 1.2511
\ea
\right.
\label{eq:15}
\ee
and $\gamma$, the nuclear surface energy constant, given by
\be
\gamma =0.9517\left[1-1.7826\left(\frac{N-Z}{A} \right)^{2}
\right] MeV fm^{-2}.
\label{eq:17}
\ee
The overlap or separation distance, in units of b, between the two
colliding surfaces
$$s(T)={{R-C_t(T)}\over b(T)}$$
with the temperature dependence of radii $R_i$, used in $C_i$, taken from Ref. \cite{davidson94} as
\be
R_i(T)=1.07(1+0.01T)A_{i}^{1\over 3}
\label{eq:18}
\ee
and surface width \cite{royer92},
\be
b(T)=0.99(1+0.009T^2).
\label{eq:19}
\ee
Here the proximity potential $V_P$ of Ref. \cite{blocki77} is extended to include the T-dependence. Similarly,
\be
V_{\ell}(T)={{\hbar ^2\ell (\ell +1)}\over {2{I(T)}}},
\label{eq:20}
\ee
where, in the complete sticking limit,
\be
I(T)={I}_{S}(T)=\mu R_a^2+{2\over 5}A_HmC_1^2+{2\over 5}A_LmC_2^2.
\label{eq:21}
\ee
Here $\mu =[A_HA_L/(A_H+A_L)]m={1\over 4}Am(1-\eta ^2)$ is the reduced mass, with m as the nucleon mass. Note that this limit is defined for
the separation distance $\Delta R$, or $\overline{\Delta R}$, to be within the range of nuclear proximity ($<$2 fm).

Finally, the $\ell_{max}$-value in Eq. (\ref{eq:3}) is the critical
$\ell$-value, in terms of the bombarding energy $E_{c.m.}$, the reduced mass $\mu$ and the first turning point $R_a$ of the entrance channel
$\eta _{in}$, given by
\be
\ell_c=R_a\sqrt{2\mu [E_{c.m.}-V(R_a,\eta _{in},\ell =0)]}/\hbar,
\label{eq:22}
\ee
or, alternatively, it could be fixed for the vanishing of fusion barrier of the incoming channel, called $\ell_{fus}$ or else the
$\ell$-value where the light-particle cross section 
$\sigma_{LP} \rightarrow 0$. This, however, could also be taken as a variable parameter \cite{sanders89,sanders87}.

The mass parameters $B_{\eta \eta}(\eta )$, representing the kinetic
energy part in Eq. (\ref{eq:4}), are the smooth classical hydrodynamical masses \cite{kroeger80}, since we are dealing here with temperatures where the shell effects are almost completely washed out.

The P in Eq. (\ref{eq:3}) is the WKB integral,
\be
P=exp[-{2\over {\hbar}}{{\int }_{R_a}^{R_b}\{ 2\mu
[V(R)-Q_{eff}]\} ^{1/2} dR}],
\label{eq:23}
\ee
solved analytically \cite{malik89}, with $R_b$ as the second turning
point (see Fig. 1) satisfying
\be
V(R_a)=V(R_b)=Q_{eff}=TKE(T).
\label{eq:24}
\ee
Note that, since we do not know how to add the $\ell$-effects in binding energies, the $\ell$-dependence of $R_a$ is defined by
\be
V(R_a)=Q_{eff}(T,\ell=0),
\label{eq:24a}
\ee
i.e. R$_a$ is the same for all $\ell$-values, given by the above equation, and that $Q_{eff}(T,\ell)=V(R_a,\ell)$. Then, using
(\ref{eq:24}), $R_b(\ell)$ is given by the $\ell$-dependent scattering
potentials, at fixed T,
\be
V(R,T,\ell)=Z_HZ_Le^2/R(T)+V_P(T)+V_{\ell}(T),
\label{eq:25}
\ee
which is normalised to the exit channel binding energy. This means that
all energies are measured {\it w.r.t.} $B_L(T)+B_H(T)$. Such a potential is
illustrated in Fig. 1 for the $^{12}C$ decay of $^{56}$Ni$^*$ at T=3.39 MeV ($E_{c.m.}$=51.6 MeV), using different $\ell$-values. The first
turning point $R_a$ is marked for 
$R=R_a=C_t+\overline{\Delta R}$-values and the second turning point
$R_b$ for only the $\ell=0$ case of $R_a=C_t+\overline{\Delta R}$. Note
that as the $\ell$-value increases, the $Q_{eff}$(=TKE)-value increases, since the decay path for all the $\ell$-values begins at $R=R_a$. For the reaction under study, at $E_{c.m.}=$51.6 MeV, the critical angular momentum $\ell_c=32 {\hbar}$ from Eq. (\ref{eq:22}) and, for the vanishing of barrier for incoming channel, $\ell_{fus}=49 {\hbar}$. It is interesting to note that this $\ell_c$ value is only slightly lower than the value of $\ell_{crit}=34 {\hbar}$ that has been extracted~\cite{sanders89,matsuse97} from the complete fusion cross section data \cite{sanders87,hinnefeld87,gutbrod73} using the sharp cutoff approximation. The corresponding values for the higher energy $E_{c.m.}$=60.5 MeV (T=3.60 MeV) case will be given later, at the appropriate place.

\section{CALCULATIONS}

The reaction $^{32}$S+$^{24}$Mg$\rightarrow ^{56}$Ni$^*$ has been studied at two incident energies ($E_{c.m.} = $51.6 and 60.5 MeV)
which correspond to 1.8 and 2.2 times the Coulomb barrier
\cite{sanders89,sanders87}. As already mentioned in the Introduction, at such energies, the fission-like IMF cross section $\sigma_{IMF}$ is about 6\% of the LP emission cross section $\sigma_{LP}$. Specifically,
${\sigma_{IMF}}/{\sigma_{LP}}={(59\pm 12) mb}/{(1050\pm 50) mb}$
at $E_{c.m.}=$60.8 MeV \cite{sanders87}. In the following, however, we
choose to use the data of Ref. \cite{sanders89}, and the above noted
energy of Ref. \cite{sanders87} is close to one of the energies of the chosen data. In the experiment of Ref. \cite{sanders89}, for the fission-like decay of $^{56}Ni^*$, a complete mass spectrum of IMFs is measured, beginning with mass $A_L$=12 fragment. Later on the measurements of the IMF mass spectrum have been extended at $E_{lab}$ = 130 MeV (equivalently, $E_{c.m.}$ = 55.7 MeV) in order to include the $^8Be$ fragment yields \cite{thummerer01}. In this experiment
~\cite{thummerer01}, an enhanced emission yield for $^8Be$ by a
factor of 1.5 to 1.8, over the two $\alpha$-particles, is observed, which in the Extended Hauser-Feshbach method (EHFM) \cite{matsuse97} is shown to be related to an increased deformation of the heavier fragment $^{48}Cr$. More recently, ternary events from a conjectured hyperdeformed $^{56}$Ni CN have been observed at $E_{lab}$ = 165 MeV (equivalently, $E_{c.m.}$ = 70.7 MeV) \cite{oertzen04}. The two sets of data (Ref.~\cite{thummerer01} and ~\cite{oertzen04}) are consistent with the strong deformation effects found in $^{56}$Ni, as is also populated by the $^{28}$Si+$^{28}$Si fusion-evaporation reaction
\cite{bhatta99,bhatta02}.


Figure 2 gives for $^{56}$Ni$^*$ the mass fragmentation potentials
$V(A)$ at different $\ell$-values, for the fixed T=3.39 MeV and
$R=R_a=C_t+1.28$ fm. The R-value is chosen for the best fit to the IMF cross section data (see below). Two interesting results are apparent
in this graph: (i) Because of T-dependent $V_{LDM}$, the non-$\alpha$, Z=N even-A, fragments also appear at minima which are in addition to the preferred $\alpha$-nucleus structure present in the macroscopic liquid drop energy due to the "Wigner term". The shell corrections $\delta U$ are already nearly zero at these temperatures. In other words, with the addition of temperature in the potential, not only the shell structure effects vanish but also there is no explicitly preferred $\alpha$-nucleus structure left. It is worth noting that the same behavior is also known for fission calculations 
\cite{sanders99}, based upon either the saddle-point picture 
\cite{charity88,moretto75,sanders89} or the scission-point picture \cite{matsuse97}. (ii) The structure in the fragmentation potential (the positions of minima and maxima) is independent of the $\ell$-value, though very important effects of symmetric or asymmetric mass division are present here in Fig. 2. Apparently, the favored (lower in energy) asymmetric mass distribution at zero and smaller $\ell$-values go over to the symmetric one for partial waves with angular momenta $\ell$ near the $\ell_c$-value. In particular, at lower $\ell$-values the light particles (plus the corresponding heavy fragments) are strongly favored, over the heavier fragments (IMFs), but this situation gets reversed at or near the $\ell_c$-value. As we shall see in the
following, these results have important consequences for the relative
contributions of LP and IMF emission at different $\ell$'s to the total
decay cross section. It may be relevant to mention here that we already know from the experiments of Beckerman {\it et al.} \cite{beckerman81} that the emission of IMFs starts only beyond a certain energy (and hence beyond a certain angular momentum) and that, for lower energies, only LP emission occurs which gives the complete fusion cross section. This result is already given by the DCM worked out in s-wave ($\ell$=0) approximation \cite{gupta02,gupta03,bala03}.


Figure 3 shows the calculated preformation factors $P_0(\ell,A_L)$, the
penetration probability $P(\ell,A_L)$ and the cross section
$\sigma_{\ell}(A_L)$, with $\ell$ summed over $\ell_{max}=\ell_c$, for
use of the fragmentation potentials of Fig. 2. Only the energetically
most favored mass fragments are considered for both the LPs and IMFs.
First of all, two important results can be drawn for $P_0(A_L)$ (shown as solid thin line in Fig. 3): (i) $P_0(A_L)$ is a strongly oscillating function with maxima only at Z=N, even-A fragments; (ii) the preformation yields are large for light-particles and the asymmetric fragments. In other words, in agreement with experiments 
\cite{sanders89,sanders87}, an asymmetric mass distribution of IMFs is favored by the preformation factors. On the other hand, the penetrability $P(A_L)$ (dashed line) is almost a monotonically decreasing function, with P being relatively small for symmetric fragments. Thus, $P(A_L)$ also support the asymmetric mass
distribution. The total cross section
$\sigma(A_L)=\sum_{\ell=0}^{\ell_c}\sigma_{\ell}(A_L)$ for
each fragment (solid thick line ) is a combined effect of these two terms, to be discussed below in comparison with experiments. Here we notice that its behavior is given more by $P_0(A_L)$ than by $P(A_L)$. In the following, we first study the variation of these quantities with angular momentum $\ell$.

%

Figure 4 shows the variation of $P_0$ with $\ell$ for the energetically
favored LPs ($A\le 4$) and the even-A, N=Z IMFs (the contribution of the energetically unfavored odd-A IMFs is small at all $\ell$'s). The
maximum $\ell$-value, $\ell_{max}$, is taken to be eqaul to the 
$\ell_c$-value. We notice that, whereas $P_0$ decreases for LPs with an increase of $\ell$, it increases for IMFs as $\ell$ increases and then starts to decrease at a large $\ell$-value. $P_0(\ell)$ for $^4He$ behaves like the LPs and that the behaviour of all LPs (A$\le $4) is different from that of the IMFs (A$>$4). Also, for heavier IMFs
(A$\ge $16), $P_0$ is almost zero for $\ell \le 18 \hbar$. Furthermore,
in Fig. 5, the $P$'s for LPs are also large, rather the largest ($P=1$
for proton emission since there is no barrier at all $\ell$'s), but the
same for $\ell \le 18 \hbar$ is nearly zero for all IMFs. Thus, for the
penetrability P also, the behavior of LPs is different from those of the IMFs. This result for the cross sections means that the lower $\ell$ values contribute mostly to the LP cross section $\sigma_{LP}$ and that the higher $\ell$ values ($\ell >18 \hbar$) contribute to the fission-like IMFs production cross section $\sigma_{IMF}$. This is illustrated in Fig. 6, where $\sigma_{LP}$, $\sigma_{IMF}$ and the total cross section $$\sigma_{Total}=\sigma_{LP}+\sigma_{IMF}$$
are plotted for each $\ell$. (The values of the three cross sections
for $\ell_{max}=\ell_c$ are also given in the brackets of the legend).
We notice that $\sigma_{LP}$ is already zero for $\ell\ge 31\hbar$ (a value close to $\ell_c$), which means that not only the two processes of LP and IMF emissions get separated at $\ell\approx 18\hbar$, but also the decay process stops at $\ell=\ell_c$-value (at least for the LPs in the present calculations). The IMF cross section starts to contribute only for $\ell >18 \hbar$ and is maximum at $\ell=\ell_c$. Furthermore, if the $\ell$ coordinate is extended up to, say, 
$\ell_{fus}$ (not shown in the figure), the contribution to LP emission remains zero but the IMF cross section goes on increasing as $\ell$ increases. This means that if the decay process continues beyond 
$\ell\approx \ell_c$, only $\sigma_{IMF}$ contributes to 
$\sigma_{Total}$. In other words, in DCM, $\ell_{max}=\ell_c$
seems to be an automatic choice, fixed by the initial conditions of the
experiment, as in Eq. (\ref{eq:22}). Alternatively one could choose $\ell_{max}$ at an $\ell$-value where $\sigma_{LP} \rightarrow 0$, as is discussed below.

%

The individual contributions of IMFs are illustrated in Fig. 7, for the $^8Be$ emission and the $^{28}Si$ emission. We notice that the contributions of the lighter IMFs towards $\sigma_{IMF}$ are much larger than that for heavier IMFs. This result is consistent with the observation of an asymmetric mass distribution, which is favored. It is interesting to note that the same results (as presented in Figs. 6 and 7) can be obtained in the statistical fission model calculations for this reaction (see, for example, Fig. 14 in \cite{sanders89}). The noticeable difference is that, in the statistical fission model 
\cite{sanders89,sanders87}, $\sigma_{fission}$ 
($\equiv \sigma_{IMF}$) also reduces to zero at $\ell=\ell_c$, which is due to the chosen phase space (the sharp cutoff approximation) in that model. Another point of interest to note in DCM is that the so-called promptly emitted LPs are really not that prompt but they do have a considerable overlap with the binary-decay process (of cluster emission) for the higher $\ell$-values. This is also consistent with the statistical fission model of \cite{sanders89,sanders87}. It is true that LP emission starts early but continues along with the emission of IMFs till the decay process itself stops for $\ell>\ell_c$.

%

Before giving the quantitative comparison of $\sigma_{LP}$ and
$\sigma_{IMF}$ calculations with experimental data, we study here the role of changing $\overline{\Delta R}$-value and the nature of emitted
light-particle(s). Fig. 8 shows the results of DCM with the use of different $\overline{\Delta R}$-values for the LPs and IMFs, taking
$\overline{\Delta R}=0.41$ fm for LPs, but keeping the same
$\overline{\Delta R}$=1.28 fm (as in Fig. 6) for IMFs. We notice that
the magnitude of (total $\ell$-summed) $\sigma_{LP}$ reduces considerably (by a factor of about 2) where as $\sigma_{IMF}$ remains nearly the same (rather an increase by about 20\% can be observed). Furthermore, if we also change the proton-emission to neutron-emission, as in Fig. 9, the $\ell_c$-value remains the same, but the magnitude of $\sigma_{LP}$ reduces further to about 60\%, keeping the $\sigma_{IMF}$ almost unchanged. These results in Figs. 8 and 9 are to be compared with the respective measured values of $\sigma_{LP}$=1050 mb and $\sigma_{IMF}$=60 mb. The agreement for $\sigma_{IMF}$ can be further improved if the $\ell$-values are summed only upto the point where
$\sigma_{LP} \rightarrow 0$; then the calculated $\sigma_{IMF}$=103 or
106 mb in Fig. 8 or Fig. 9, respectively. Note that the drastic reduction of $\sigma_{LP}$ in Fig. 9 occurs because the lower 
$\ell$-values ($\ell\le 3 \hbar$) also do not contribute to the LP emission cross section. It is known for experiments that it is more difficult to evaporate a neutron than a proton, as is also shown to be energetically the case in Fig. 2. However, we do not make a search for the exact emission of LPs, since we do not attempt a one-to-one comparison with the data. On the other hand, the general success of DCM is demonstrated here for the first time to give the LP emission in a non-statistical formalism.

%

The total cross section $\sigma=\sum \sigma_{\ell}$ for decay of
$^{56}$Ni$^*$ at $E_{c.m.}=$51.6 MeV to light fragments $A_L$ (for both
the LPs and even-A, N=Z IMFs) is plotted in Fig. 10 for the case of
Fig. 8, and compared with other calculations based either on the
saddle-point transition-state model (TSM) \cite{sanders89} or on the scission-point model, the so-called EHFM scission-point model 
\cite{matsuse97}, and available experimental data 
\cite{sanders89,sanders87} for $A_L\ge 12$ IMFs. The TSM calculations for LPs are performed within the HF formalism \cite{sanders89}, and hence are the same for EHFM model \cite{matsuse97}. The EHFM calculations for IMFs are done for $A_L\ge$12 fragments and are thus joined straight from $A_L=$4 to $A_L=$12. Only even-A fragments are plotted since the IMF-spectra at $E_{c.m.}=$51.6 MeV is measured for only even-A fragments \cite{sanders89,sanders87}. For the LP emission at this energy, as already stated above, the measured (fusion or evaporation residue) cross section is available 
($\sigma_{LP}=1050\pm 100$ mb) but the separate yields for each
emitted LP is not given for a possible direct comparison between the experiment and model calculations.

In Fig. 10, we can notice that the known discontinuity at the point between $A_L=$4 and 6 in both TSM and EHFM calculations, due to the use of HF formalism for LPs ($A\le $4), is no more present in the present DCM calculations. The DCM treats both the LPs and IMFs emission in a similar manner, although the present calculations (corresponding to the case of Fig. 8) overestimate $\sigma_{LP}$ by a factor of more than two. This discrepancy in DCM is mainly due to the contribution of 
mass-one-particle, here the proton. For example, the replacement of proton with neutron (Fig. 9) yields the calculated $\sigma_{LP}$
to be in better agreement with the data. Apparently, for the calculation of the evaporation residue cross section $\sigma_{LP}$ or $\sigma_{LP}+\sigma_{IMF}$, it is very important to know exactly the contributing particles (i.e. their multiplicities) for comparisons with experiments. Note further that the HF analysis gives nearly equal cross sections for each of the four emitted particles ($A_L$=1-4), whereas a decreasing function of the light-particle mass is obtained in DCM. It will be of high interest to measure the LP's trends in the near future.

For the IMFs, in Fig. 10, the general comparison between the experimental data and DCM is of the same quality as for the TSM or EHFM, at least for $A_L\le 22$. For $A_L>22$ the TSM and EHFM predictions appear to be better. Note that in DCM calculations $\overline{\Delta R}$=0.41 fm for LPs and $\overline{\Delta R}$=1.28 fm for IMFs. However, a closer comparison of DCM calculations with experiments (Fig. 12, discussed below) favors the use of a
fragment-dependent $\overline{\Delta R}$ or the actual 
$\Delta R(\eta)$, calculated from $Q_{eff}$, presented in Fig. 11 for light mass fragments.

%

In Fig. 11, we notice that $\Delta R(\eta)$ has an oscillatory nature, if compared to the smooth variation of $C_t$ or 
$C_t+\overline{\Delta R}$ with $\eta$. The maxima in $\Delta R(\eta)$ correspond to $\alpha$-nuclei IMFs and the minima to N=Z  
non-$\alpha$-nuclei IMFs, with the odd-$A_L$ fragments lying in between. For light-particles, the $\Delta R(\eta)$ values increase almost monotonically. In any case, the division between the LPs and the IMFs is clearly evident from the variation of $\Delta R$ with $\eta$.

Figure 12 shows the DCM calculations for use of different average
$\overline{\Delta R}$-values and the actual $\Delta R(\eta)$ obtained in Fig. 11 from calculated $Q_{eff}$. These calculations are presented here only for even-A IMFs. We have also added here the DCM calculations for $\overline{\Delta R}$=1.28 (from Fig. 10). It is clear that
$\overline{\Delta R}$=1.28 fm gives the optimum fit to IMF data, though the oscillatory nature of data is almost smoothed out, particularly for the heavier IMFs. Interesting enough, this oscillatory structure of the cross section gets restored with the use of actual $\Delta R(\eta)$ obtained from calculated $Q_{eff}$, though the fit with data is now deteriorated. Apparently, an improvement in $\Delta R(\eta)$ is required. This calls for an improvement in the calculations of 
$Q_{eff}$ and hence in the ground-state and T-dependent binding energies.

The histograms in Fig. 13 show the comparisons of the absolute
IMF cross sections for the best fit ($\overline{\Delta R}$=1.28), the experimental data \cite{sanders89,sanders87} and the two alternate model-calculations of TSM \cite{sanders89} and EHFM \cite{matsuse97}, for $A_L\ge 12$ at both the available energies. Similarly, Fig. 14 shows the DCM calculated excitation functions (cross sections at different $E_{c.m.}$) for the emission of $^{12}C$ from the excited 
$^{56}Ni^*$ CN. We notice in Fig. 14 that, independent of the choice of $\overline{\Delta R}$-value (i.e. a constant or T-dependendent value), the $^{12}C$ emission cross section, $\sigma(^{12}C)$, increases as the incident energy increases and reaches a maximum around $E_{c.m.}$ = 90 MeV and then starts to decrease at higher incident energies. It is interesting to note that similar results are obtained in the HF calculations, using BUSCO code, for the emission of $^{12}C$ from 
$^{114-118}Ba^*$ CN \cite{spiral04}. From Figs. 13 and 14 it is clear that the DCM contains the required features of the experimental data, as well as of other models (EHFM~\cite{matsuse97} and 
TSM~\cite{sanders89}). A better treatment of the binding energies and missing aspects, such as the deformations of the fragments and neck formation between them, would also ameliorate the predictions of the present model. In view of this hope, in the following, we further analyze the comparisons of the DCM calculations for average 
$\overline{TKE}$s with the experimental data of Ref. \cite{sanders89}.

%

Figure 15 shows the DCM calculated average total kinetic energy
$$\overline{TKE}(A_L)={\sum_{\ell=0}^{\ell_{max}}{\sigma_{\ell}(A_L)\over \sigma (A_L)}TKE(\ell,A_L)},$$ 
compared with the experimental data for the $^{32}$S+$^{24}$Mg reaction \cite{sanders89} leading to hot $^{56}$Ni$^*$ at the two chosen energies. Here, for each fragment, the TKE for each $\ell$ is averaged over its corresponding production cross section $\sigma_{\ell}$ 
{\it w.r.t.} the total cross section $\sigma (A_L)=\sum_{\ell=0}^{\ell_{max}}\sigma_{\ell}(A_L)$. We have also calculated the total kinetic energy $TKE(A_L)$ for a best fit to the $\ell$-value. Apparently, the comparisons with data are reasonably good for both the calculations and it is difficult to distinguish between the calculated $TKE$ and $\overline{TKE}$. The maximum $\ell$-value is nearly the same in both case. However, it is not clear why this maximum $\ell$-value is much less than the $\ell_c$-value. The simple model-dependence used for handling the deformations of the fragments and neck formation between them need further improvements. 

\section{Summary}

We have further developed the DCM for the decay of a hot and rotating CN, formed in light heavy-ion reactions, into multiple LP evaporation and IMF emission. The LP emission (evaporation residue) cross section $\sigma_{LP}$, constitutes the CN fusion cross section 
$\sigma_{fusion}$ for a negligible emission of IMFs, since 
$\sigma_{fusion}=\sigma_{LP}+\sigma_{IMF}$, also referred to as $\sigma_{Total}$. The statistical equilibrated CN evaporation process, successfull for the emission of LPs, could not explain the IMF emission. Alternatively, the IMF emission alone could be understood as the statistical fission process in the saddle-point~\cite{sanders89} or scission-point model \cite{matsuse97}. On the other hand, in DCM, both the LPs and IMFs are treated identically as the dynamical collective mass motion of preformed fragments or clusters through a barrier, i.e. quantum mechanical tunnelling of clusters that are considered pre-born with different probabilities before they actually penetrate the barrier. Thus, the cluster preformation probabilities contain the structure effects of the CN, that are found to be important in the description of the measured excitation functions of large-angle elastic and inelastic scattering yields in the experiments under study.

The DCM is worked out in terms of only one parameter, the neck-length
parameter, that depends on the total kinetic energy of the fragments TKE(T) at the given temperature T of the CN, which itself is defined for the first time in terms of the binding energies of the emitted fragments in their ground-states and the binding energy of the
hot CN. The hot CN is considered to achieve its ground-state by giving away its extra binding energy to the emitted LPs, which is shown to leave the emitted IMFs in their respective ground states with total kinetic energy TKE(T=0). The remaining (excitation) energy TXE(T)
must go in to the emission of secondary light-particles from the
IMFs which are otherwise already in their ground-states in the radial motion. Such an emission of secondary light-particles is not included here in the DCM so-far; rather the model predictions are compared with the primary IMF experimental data \cite{sanders89}, corrected for such an emission.

The DCM is applied here to the decay of $^{56}Ni^*$, formed in the
$^{32}$S+$^{24}$Mg reaction at two incident energies $E_{c.m.}$= 51.6
and 60.5 MeV, where both the LP cross section and IMF spectra, as well as the total average kinetic energy ($\overline{TKE}$) for only the favored $\alpha$-nucleus fragments, are measured \cite{sanders89}. The interesting result of DCM is that both the preformation factors and penetrabilities, as a function of angular momentum, behave differently for the LPs and the IMFs. In other words, there is an explicit division at mass-four fragment between the LPs and IMFs with $^4He$ belonging clearly to the LP regime. The preformation factor is shown to contribute more to the observed behaviour of IMF cross section 
$\sigma_{IMF}$, which can be compared with the experimental data
reasonably well, favoring an asymmetric distribution. Furthermore, the
variation of both $\sigma_{LP}$ and $\sigma_{IMF}$ with angular momentum, as well as the individual contributions of IMFs to 
$\sigma_{IMF}$, and the excitation functions of the emitted IMFs, match exactly the predictions of the statistical fission model, and the HF analysis. Since, unlike fission models, the DCM does not depend on the chosen phase space it has the advantage that the $\ell_{max}$-value is fixed by the initial conditions of the experiment via $\ell_c$, rather than by the available phase space. This distinguishing feature is evident in $\sigma_{IMF}$ not going to zero when $\sigma_{LP}$ goes to zero at $\ell_{max}=\ell_c$. The comparison of $\sigma_{LP}$, however, depends strongly on the type of particles involved and their multiplicities, as expected. The calculated $\overline{TKE}$ also reproduces the experimental data, though at an $\ell<\ell_c$-value, which has perhaps to do with the way the deformations of the fragments are included here simply through the same neck-length parameter that accounts for the temperature effects. The model is being improved both for the neglected deformation effects and neck-formation between them as well as the binding energies used to calculate this neck-length
parameter.

\begin{acknowledgments}

The financial support from the Department of Science and Technology (DST), and the Department of Atomic Energy (DAE), Govt. of India, in terms of research projects, is gratefully acknowledged. One of us (C.B.) would like to thank J.P. Wieleczko and W. von Oertzen for interesting discussions on different aspects of the dynamical cluster-decay model.
\end{acknowledgments}

\newpage
\par\noindent
{\bf Figure Captions:}
\begin{description}
\item{Fig. 1} {The temperature and angular momentum dependent scattering potentials, illustrated for 
$^{56}Ni^*\rightarrow ^{12}C+^{44}Ti$
at T = 3.39 MeV (equivalently, $E_{c.m.}$ = 51.6 MeV). The potential for each $\ell$ is calculated by using 
$V(R,T,\ell) = E_c(T)+V_P(T)+V_{\ell}(T)$, normalized to exit channel T-dependent binding energies $B_L(T)+B_H(T)$, each defined as 
$B(T) = V_{LDM}(T)+\delta U(T)$. The decay path, defined by 
$V(R_a,\ell) = Q_{eff}(T,\ell)$ for each $\ell$, is shown to begin at $R_a = C_t+\overline{\Delta R}$ for $\ell = 0$ case, where 
$\overline{\Delta R}$ is the average over $\eta$ of the different neck-length parameters $\Delta R(=R_a-C_t)$ calculated for
$V(R_a) = Q_{eff}(T,\ell = 0)$ for all possible fragmentations. The
critical angular momentum $\ell _c$-value is determined from 
Eq. (\ref{eq:22}). For other details, see text.}
\item{Fig. 2} {The fragmentation potentials $V(A)$ for the compound system $^{56}$Ni$^*$ formed in the reaction $^{32}$S+$^{24}$Mg at
$E_{c.m.}=$51.6 MeV, calculated for different $\ell$-values at a fixed
$T$=3.39 MeV (corresponding to $E_{c.m.}=$51.6 MeV) and 
$R_a=C_t+1.28$ fm. The fragmentation potential
$V(A)=V_{LDM}(T)+\delta U(T)+E_c(T)+V_P(T)+V_{\ell}(T)$
for fixed $R=R_a$. The $\ell _c$-value is as per Eq. (\ref{eq:22}).}
\item{Fig. 3} {The fragment preformation factor $P_0(\ell,A_L)$, the
penetrability $P(\ell,A_L)$ and the decay cross section
$\sigma_{\ell}(A_L)$, with $\ell$ summed over
$\ell_{max}=\ell_c=32 {\hbar}$ calculated on Eq. (\ref{eq:22}), for the decay of $^{56}$Ni$^*$ to various complex fragments (both LPs and IMFs), using the fragmentation potentials in Fig. 2 based on DCM. }
\item{Fig. 4} {The variation of $P_0$ with $\ell$, for both the LPs (dashed lines) and even-A, N=Z IMFs (solid lines), calculated on DCM for the compound system $^{56}Ni^*$, using the fragmentation potentials in Fig. 2. For proton, the calculated $P_0$ values are three times of what are shown in the figure.}
\item{Fig. 5} {The variation of P with $\ell$, for both the LPs (dashed lines) and even-A, N=Z IMFs (solid lines), calculated on DCM for the compound system $^{56}Ni^*$, using the scattering potentials as in Fig. 1. For proton, there is no barrier at any $\ell$-value and, hence P=1 for proton.}
\item{Fig. 6} {The variation of evaporation residue cross sections due to the
LPs $\sigma_{LP}$ (dotted line), the IMFs $\sigma_{IMF}$ (dashed line),
and their sum $\sigma_{Total}$ (solid line) with $\ell$, for
$\ell$-values up to $\ell_c$, calculated on DCM for the compound system
$^{56}$Ni$^*$ formed in $^{32}$S+$^{24}$Mg reaction at $E_{c.m.}=$ 51.6
MeV. The parameter $\overline{\Delta R}$ (=1.28 fm) is the same for
both the LPs and IMFs. The cross sections given in the brackets are obtained by summing for $\ell_{max}=\ell_c=32 \hbar$. Here the LPs consist of $^{1,2}H$ and $^{3,4}He$.}
\item{Fig. 7} {The variation of $P_0(\ell,A_L)$, $P(\ell,A_L)$ and 
$\sigma_{\ell}(A_L)$ with $\ell$, for the emission of $^8Be$ and 
$^{28}Si$ fragments from $^{56}$Ni$^*$ formed in $^{32}$S+$^{24}$Mg reaction at $E_{c.m.}=$ 51.6 MeV, calculated on DCM, as in Fig. 6. For $^{28}Si$-decay, the calculated P is ten times and, $P_0$ one tenth, of the plotted values.}
\item{Fig. 8} {The same as for Fig. 6, but for use of different 
$\overline{\Delta R}$-values for LPs and IMFs.}
\item{Fig. 9} {The same as for Fig. 8, but for the proton-emission replaced by neutron-emission.}
\item{Fig. 10} {The same as for Fig. 3, but for only the decay cross section $\sigma (A_L)$ and calculated for the case of Fig. 8. The DCM calculations are compared with two other model calculations EHFM and TSM of Refs. \cite{matsuse97} and \cite{sanders89}, respectively, and the experimental data \cite{sanders89,sanders87} for $A_L\ge 12$ IMFs.}
\item{Fig. 11} {The variation of the first turning point $R_a$ with light fragment mass $A_L$, for cases of $R_a=C_t$,
$R_a=C_t+\overline{\Delta R}$(=1.28 fm), and the actually calculated
$R_a$ from $V(R_a)=Q_{eff}(T)$, $\ell $=0, for the decay of $^{56}Ni$ at T=3.39 MeV. Note that the actually calculated $R_a$ from
$V(R_a)=Q_{eff}(T)$, $\ell $=0 can be written as 
$R_a=C_t+\Delta R(\eta)$ where $\Delta R(\eta)$ is found to be, in general, positive. For some light fragments ($A_L$=1-5,8), we use 
$\ell >$0 since calculated $Q_{eff}(T)$ is larger than the barrier for $\ell $=0 case.}
\item{Fig. 12} {The same as for Fig. 10, but for DCM alone, calculated for different average $\overline{\Delta R}$-values and the actual
$\Delta R(A_L)$ determined from $V(R_a)=Q_{eff}(T,\ell=0)$ (Fig. 11).
The DCM calculations are compared with the experimental data taken from
Ref. \cite{sanders89,sanders87}.}
\item{Fig. 13} {The histograms of the calculated IMF cross sections 
$\sigma (A_L)$ on DCM, compared with the experimental data 
\cite{sanders89,sanders87} and two other model calculations EHFM 
\cite{matsuse97} and TSM \cite{sanders89}, for even-$A_L$ IMFs at 
$E_{c.m.}=$51.6 MeV and for both the odd- and even-$A_L$
IMFs at $E_{c.m.}=$60.5 MeV. The predicted $A_L=$14 fragment cross section on DCM is very large, ten times of what is plotted here.}
\item{Fig. 14} {The DCM excitation functions i.e. the cross sections at different incident c.m. energies for emission of $^{12}C$ fragment from the excited $^{56}Ni^*$ compound nucleus, calculated for a constant and 
an arbitrary T-dependent $\overline{\Delta R}$-value. }
\item{Fig. 15}  {The measured \cite{sanders89} and DCM calculated average total kinetic energy ($\overline{TKE}$) for the reaction
$^{32}$S+$^{24}$Mg$\rightarrow ^{56}$Ni$^*\rightarrow A_L+A_H$, at two
incident energies $E_{c.m.}=$51.6 and 60.5 MeV. Also, the total kinetic energy $TKE$ for the best fit to $\ell$-value is plotted. The average 
$\overline{\Delta R}$=1.28 and 1.29 fm, respectively, for the two energies.}
\end{description}
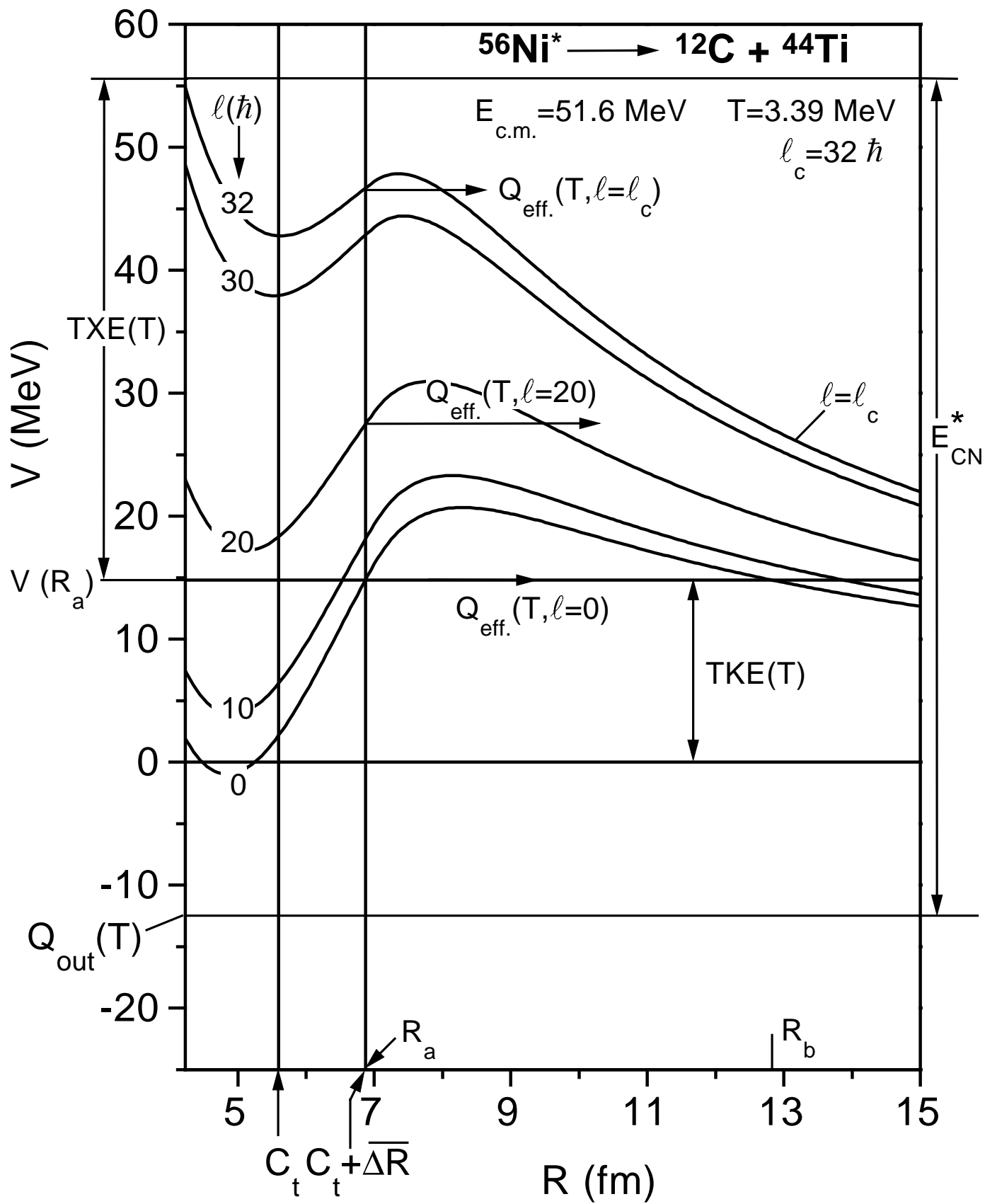
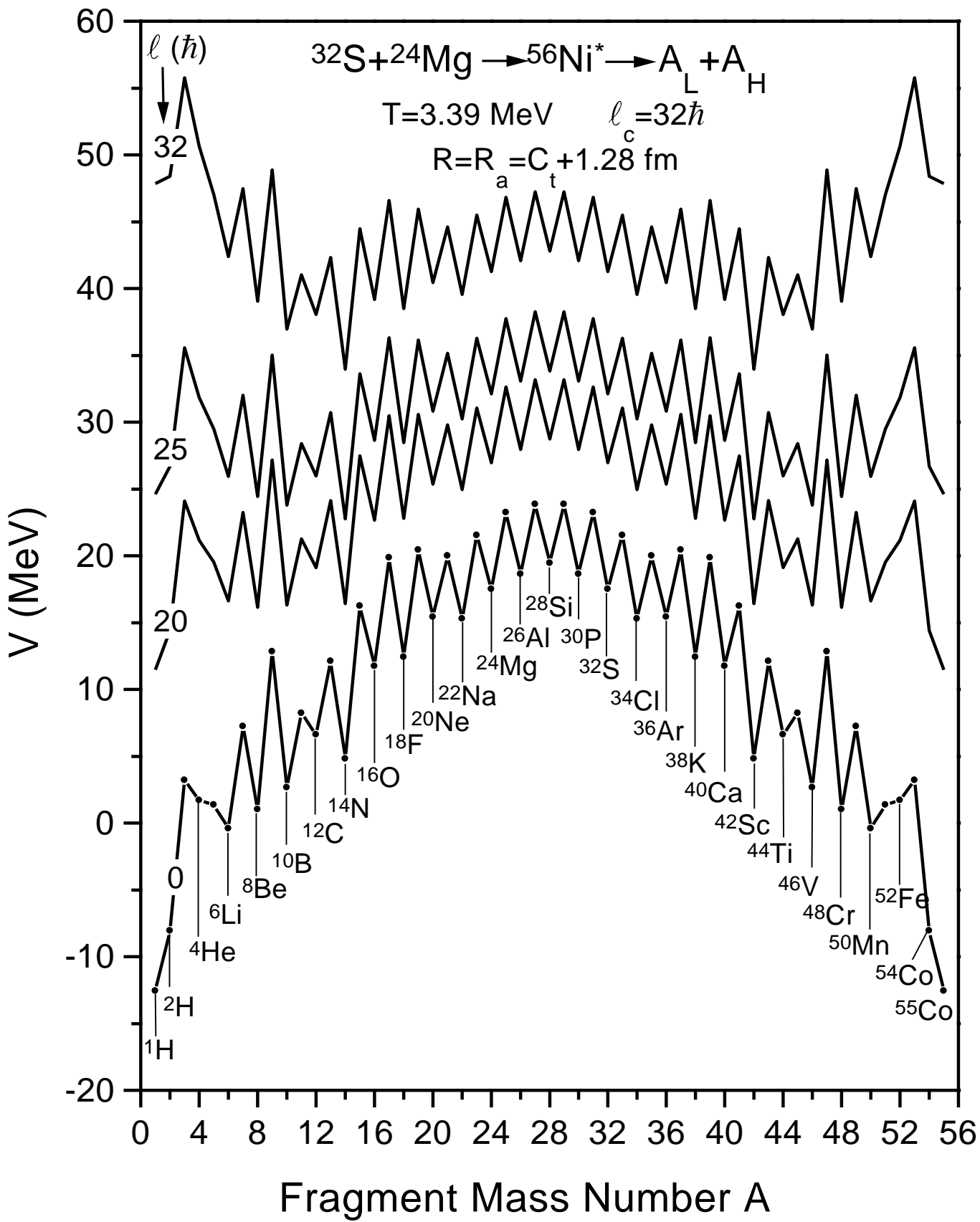
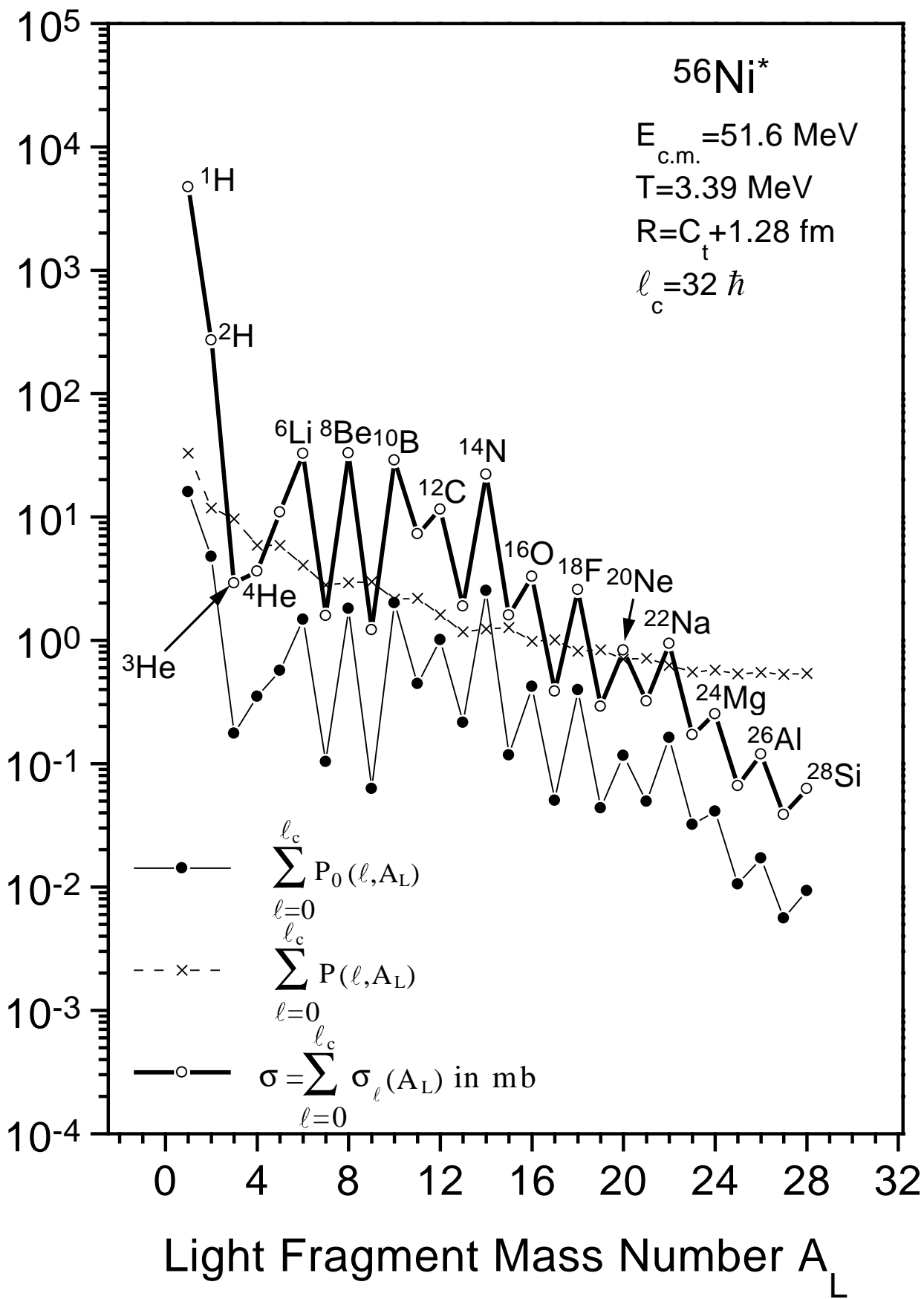
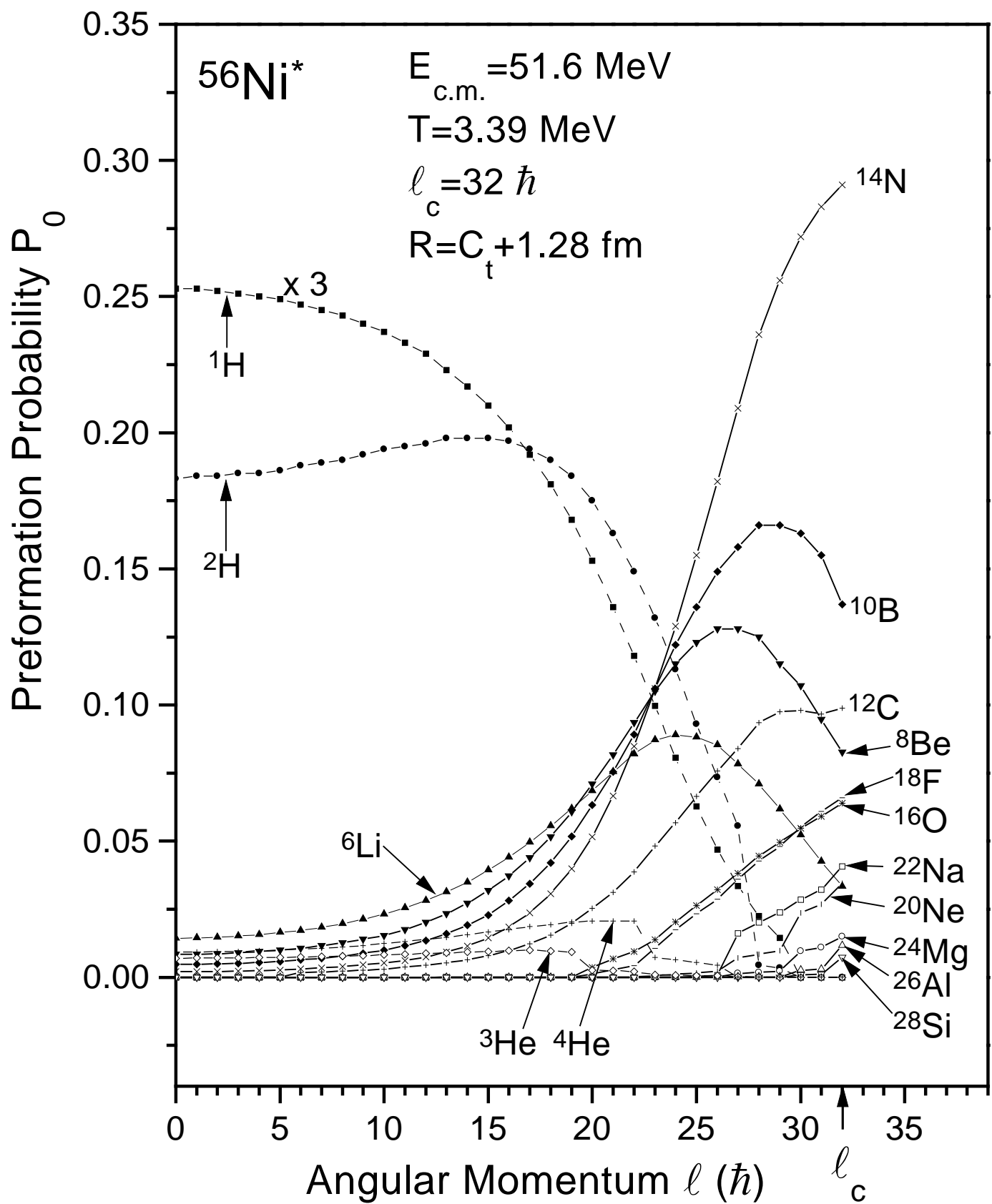
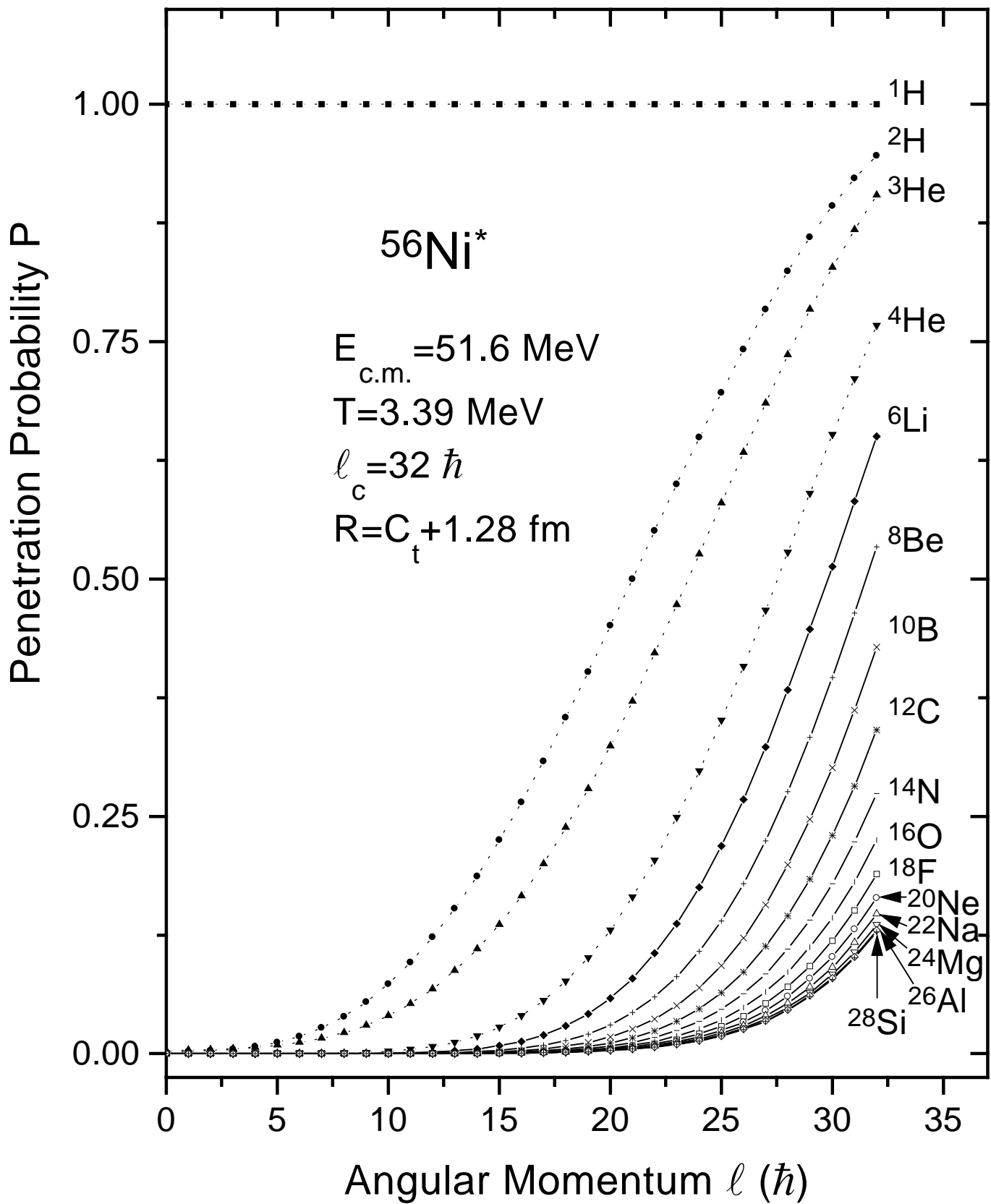
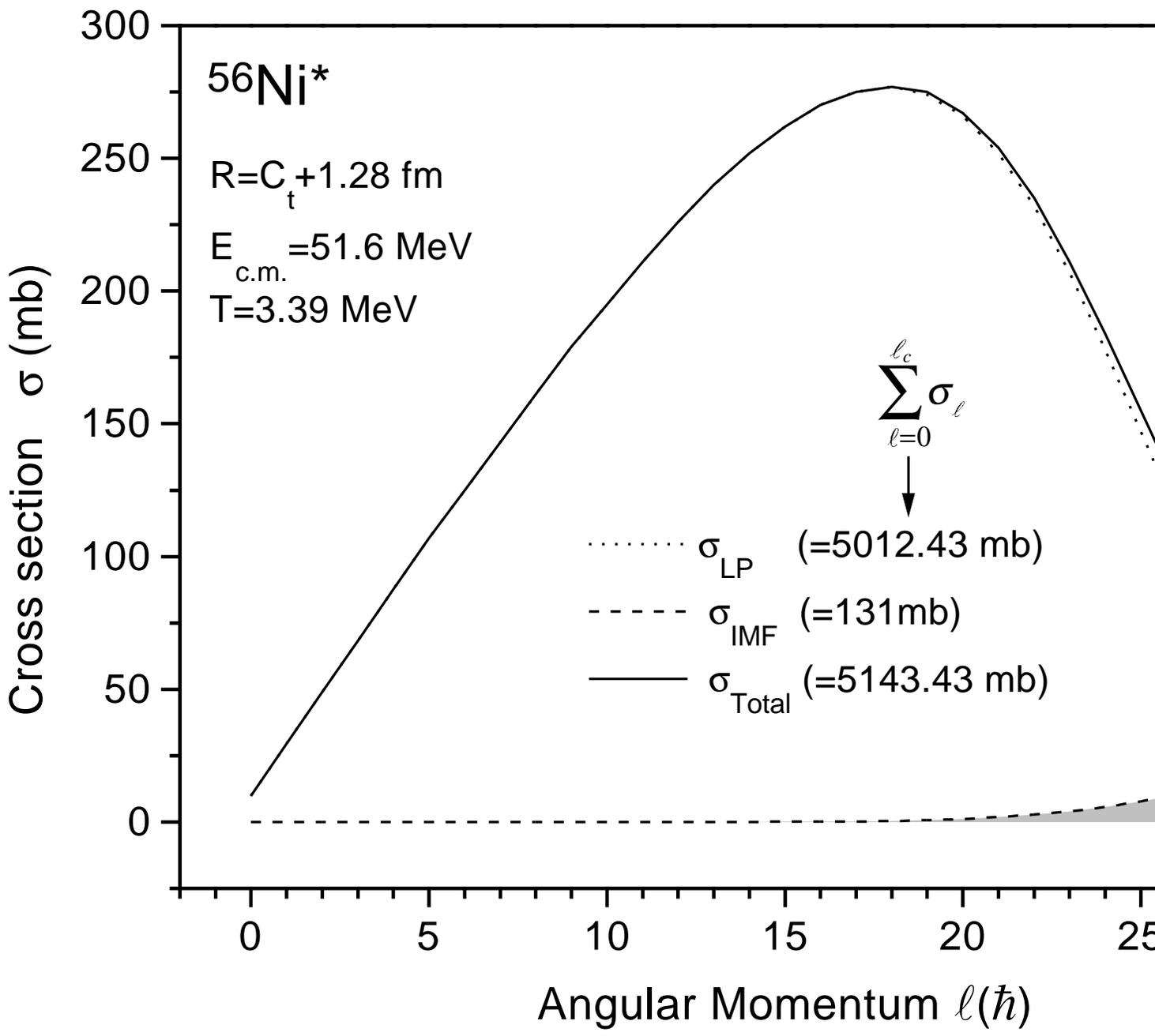
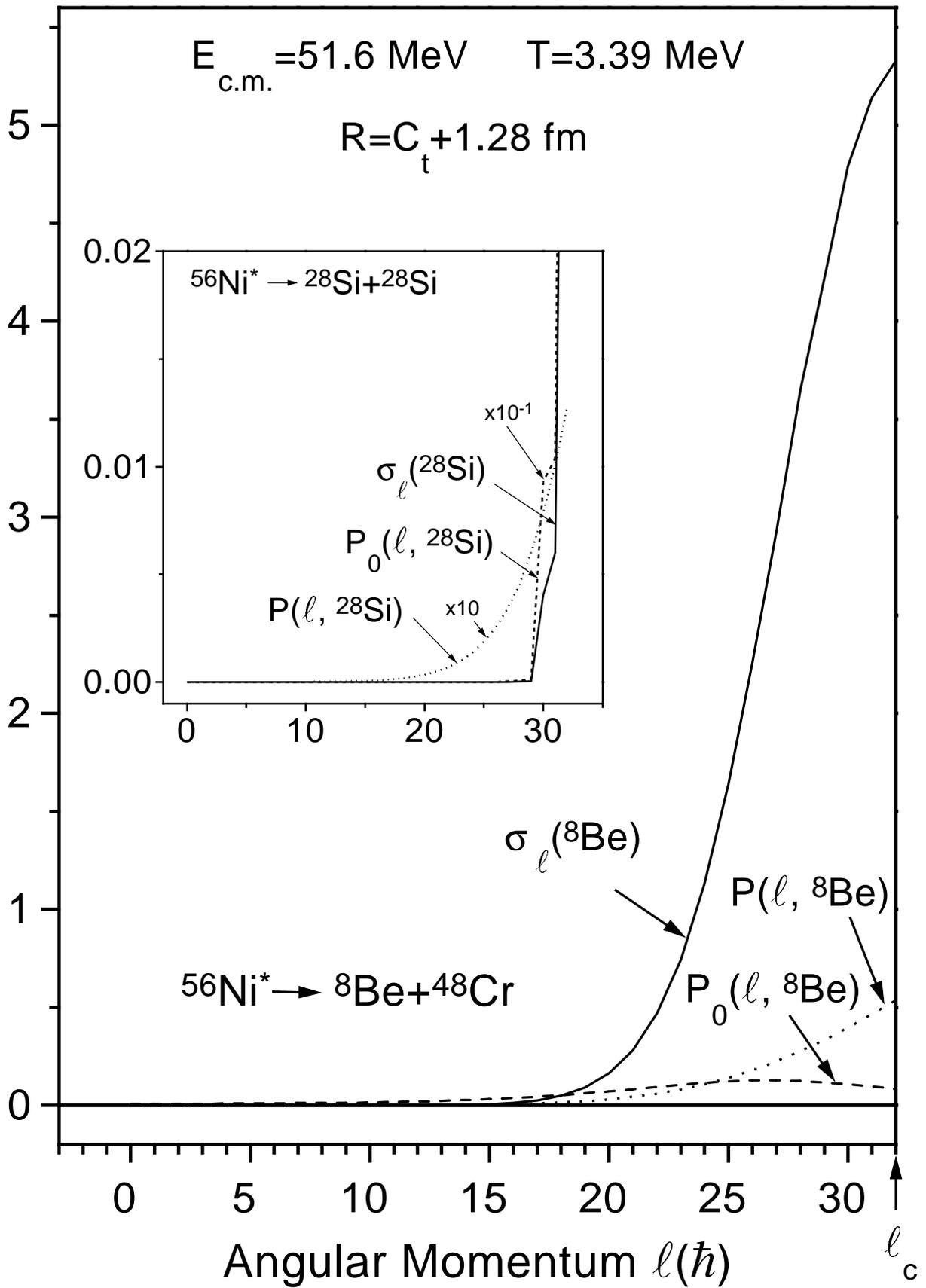
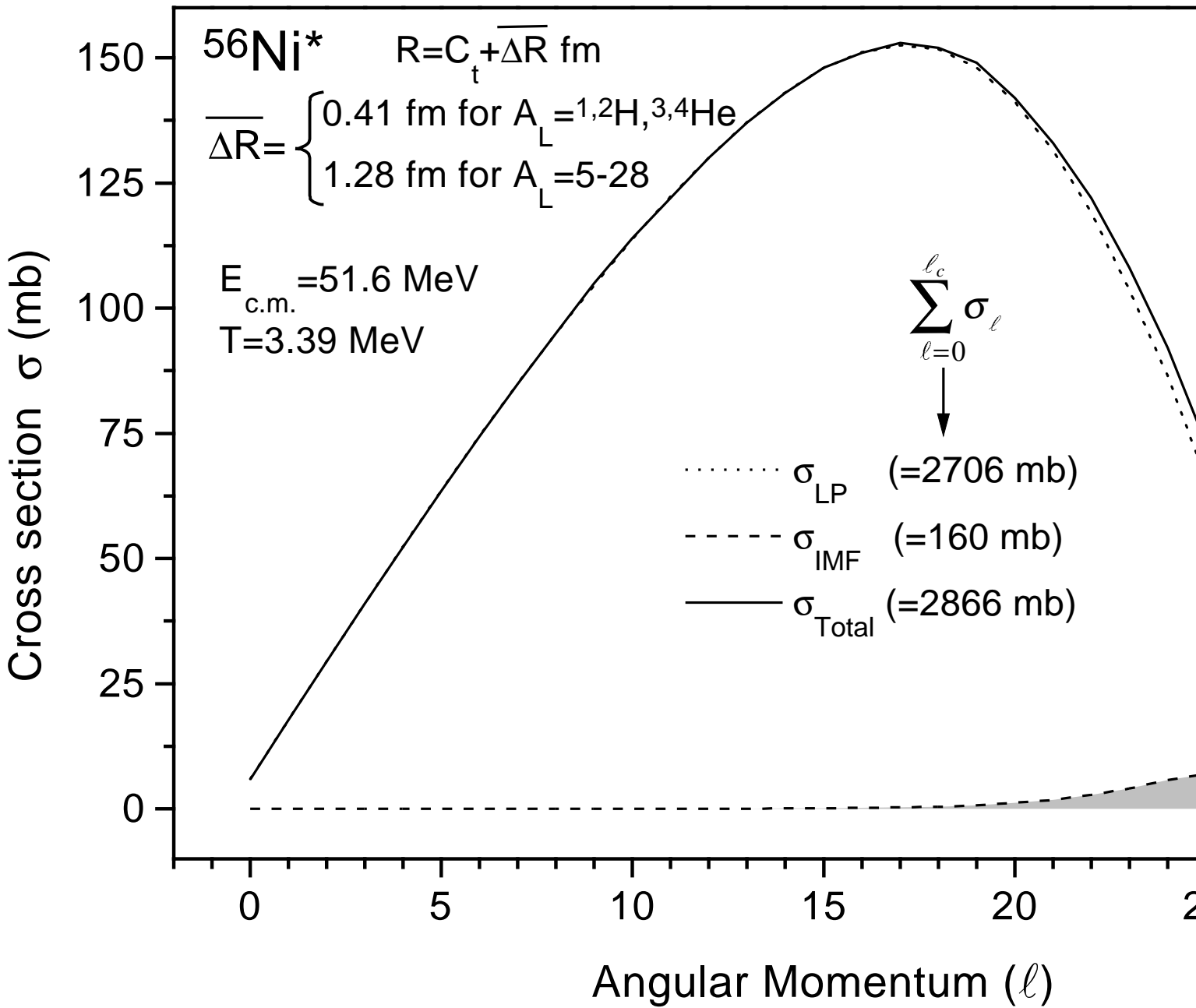
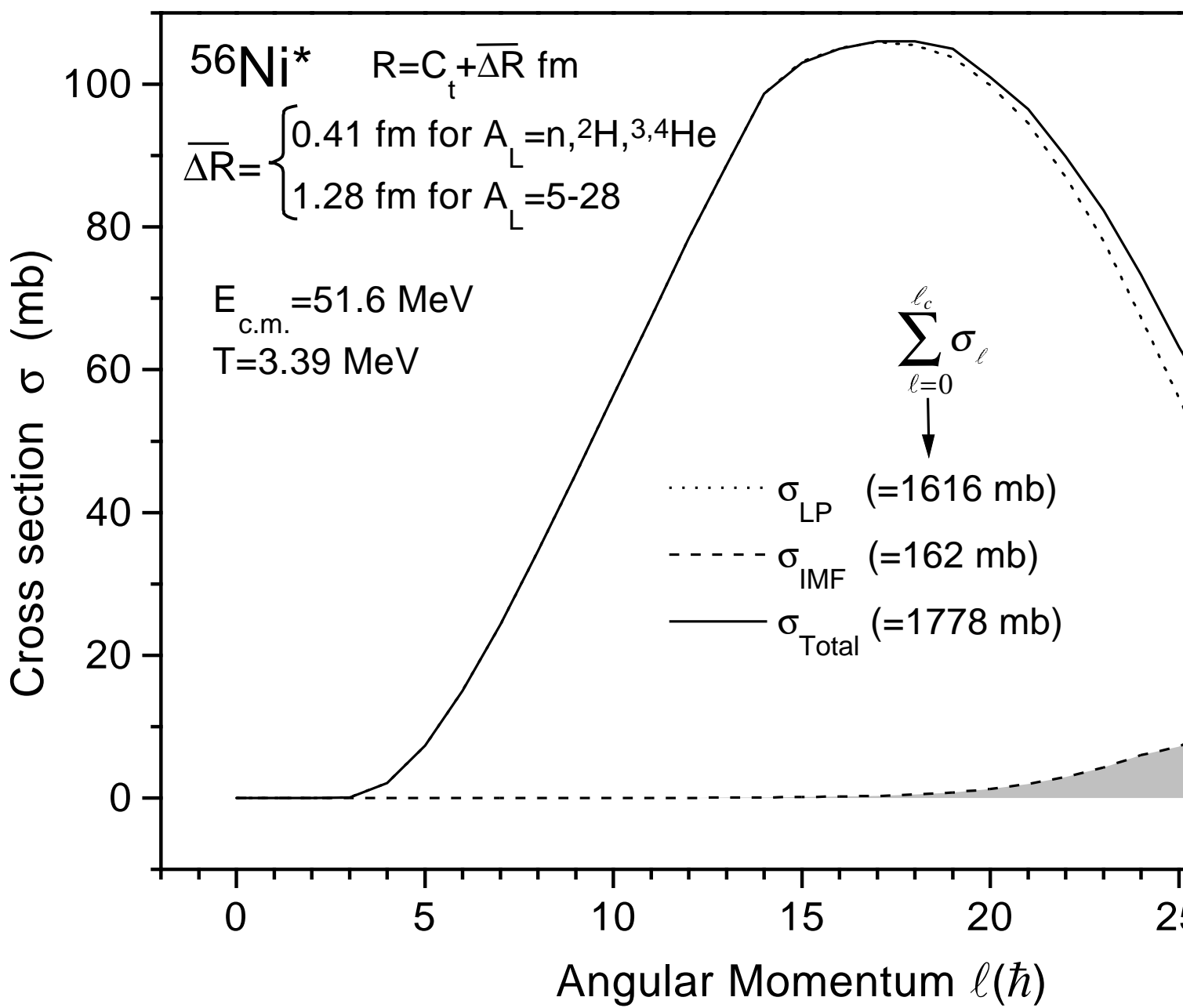
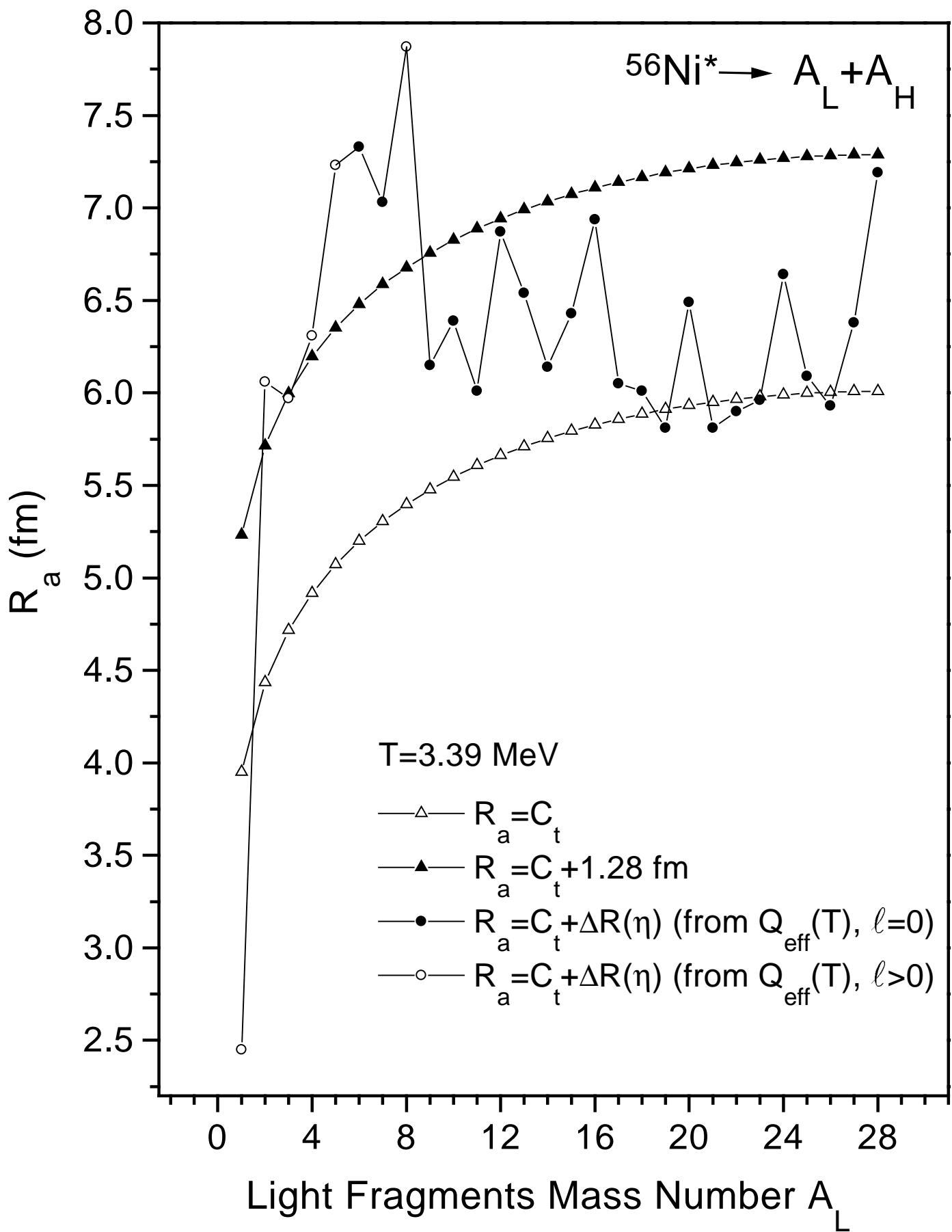
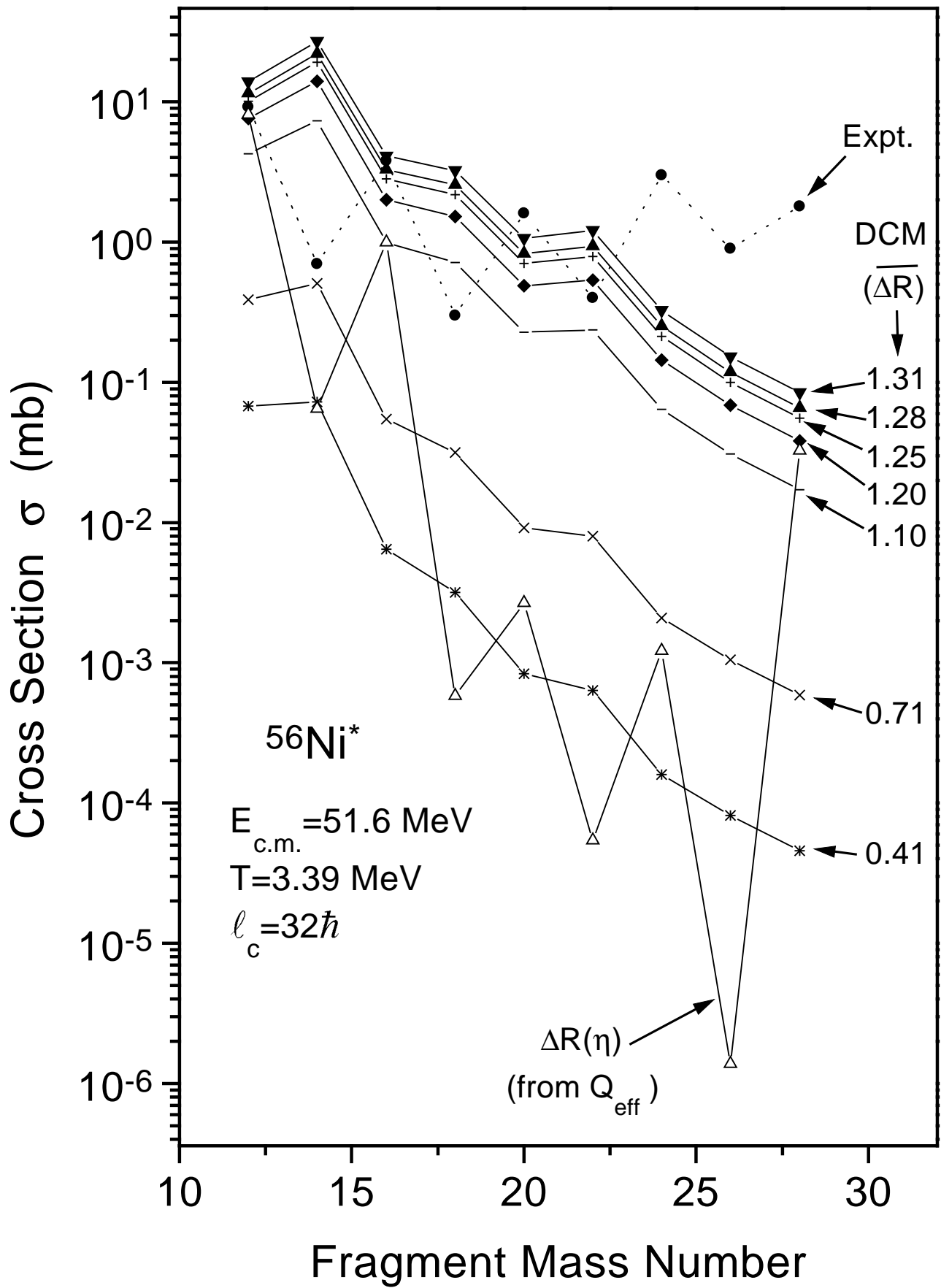
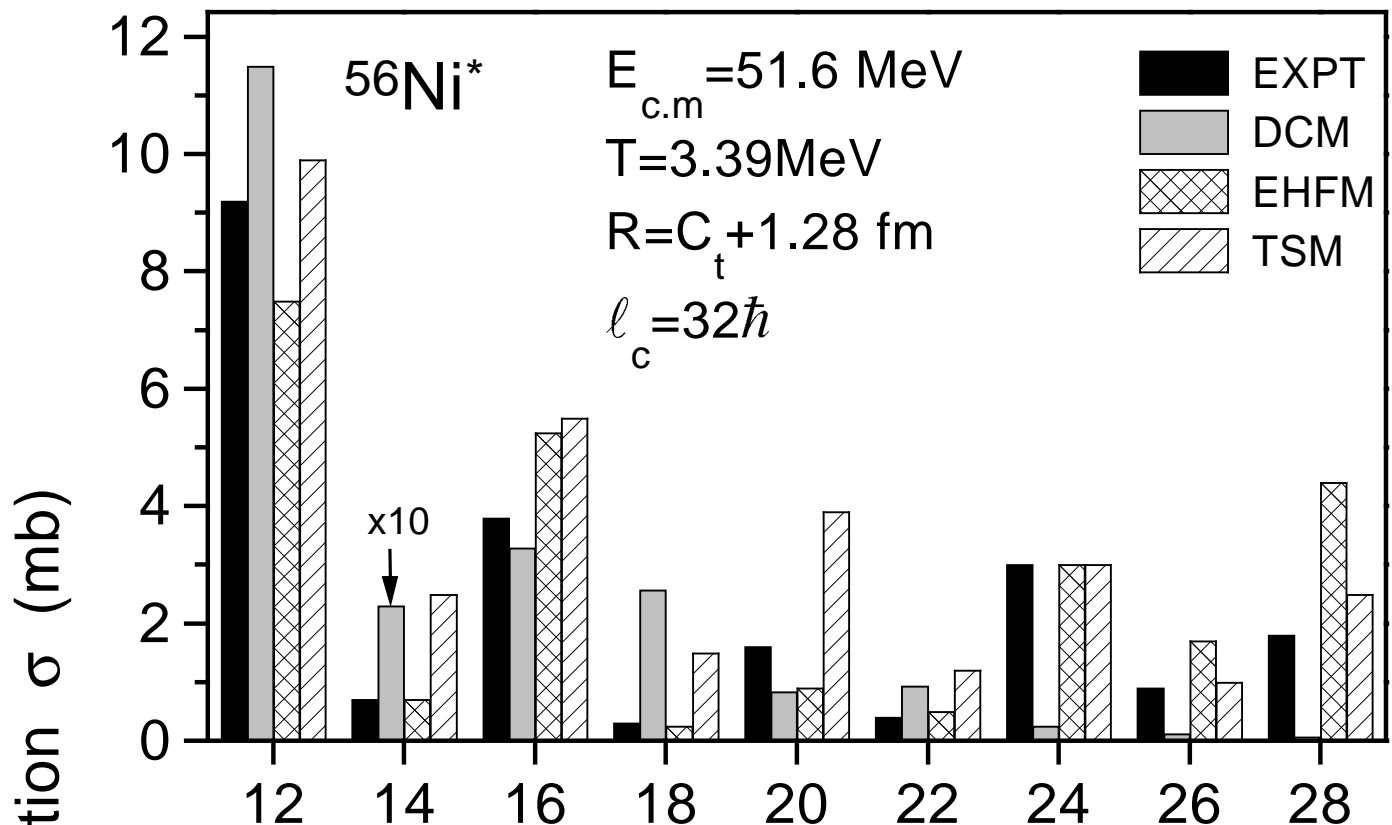
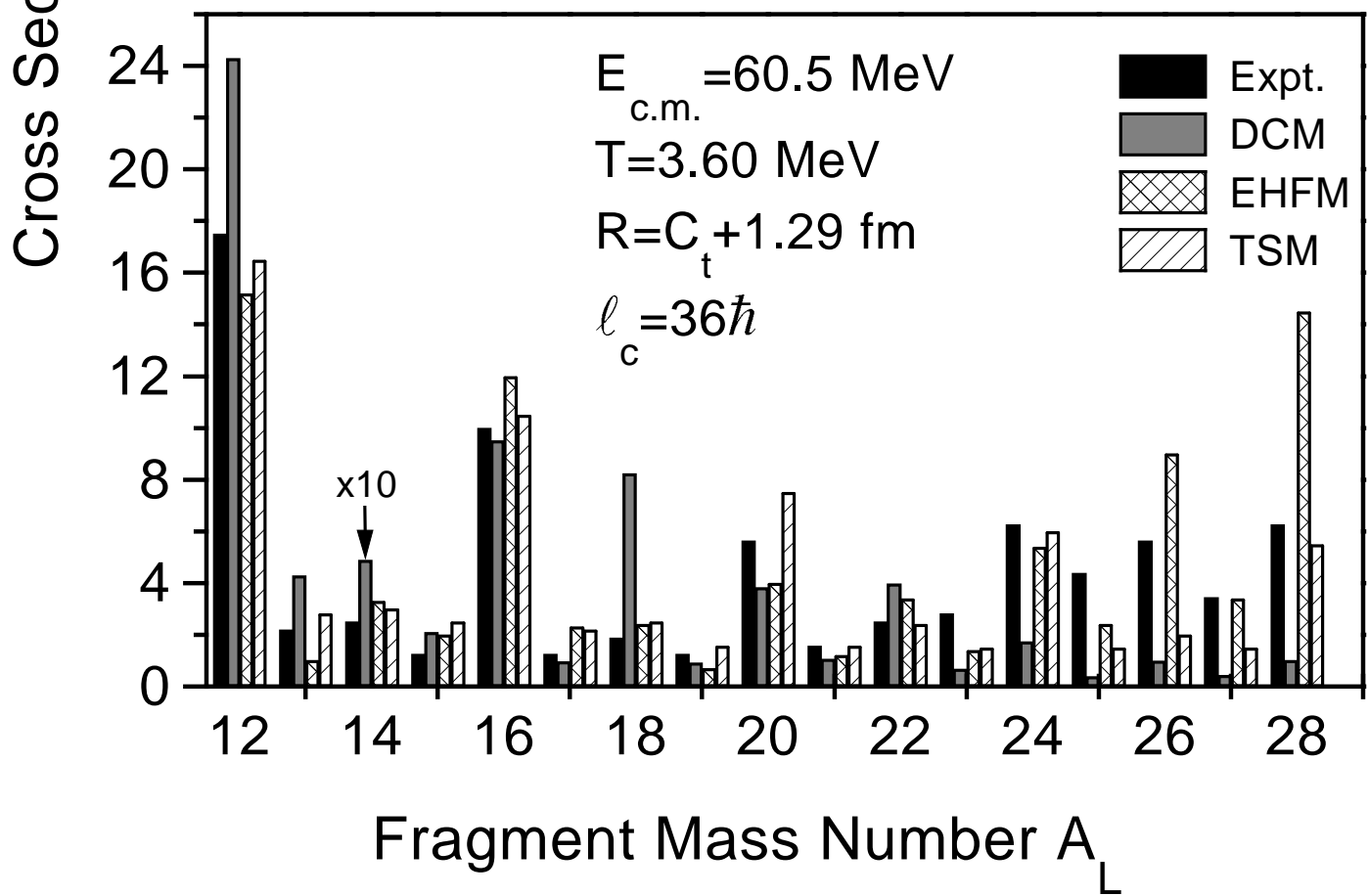
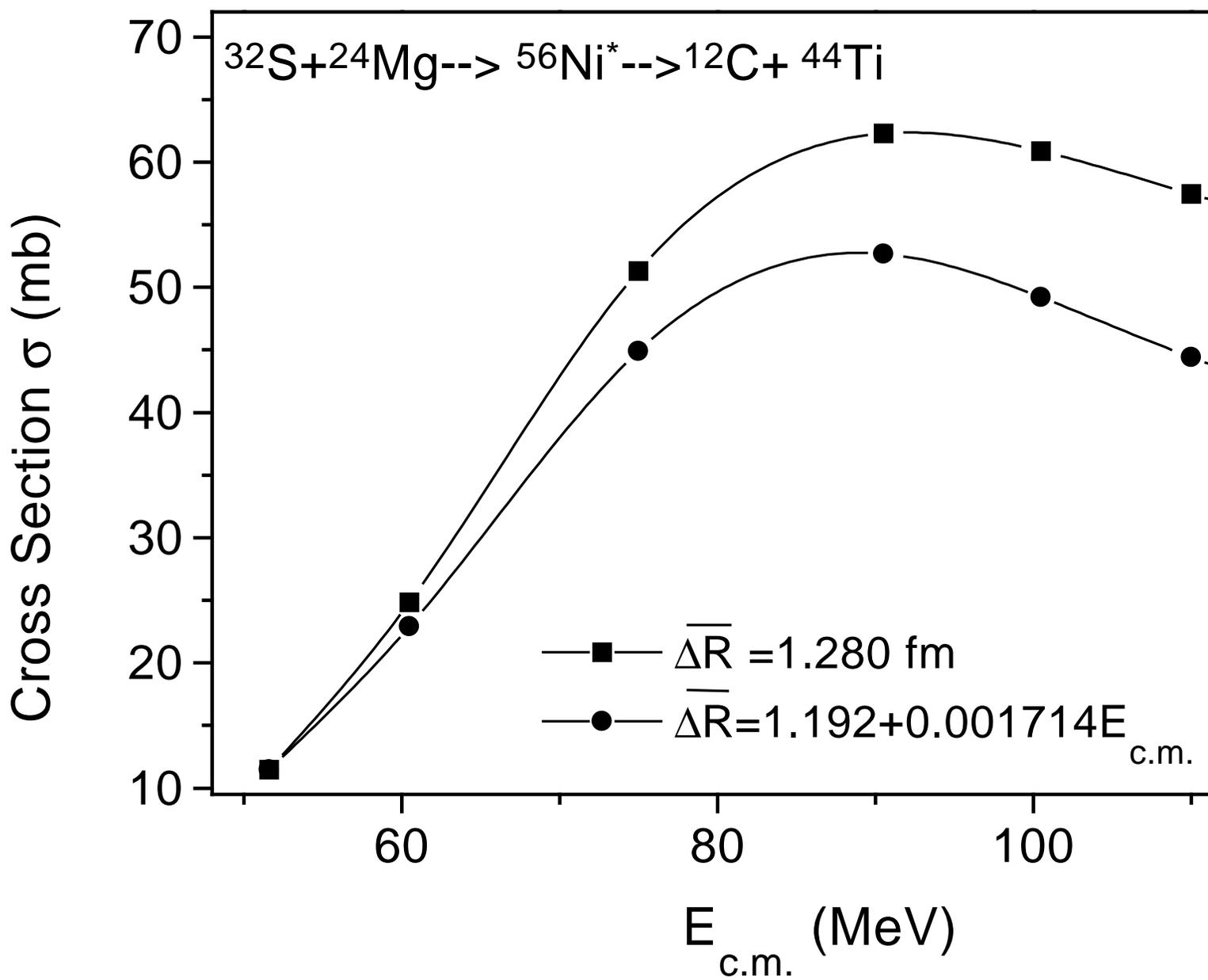
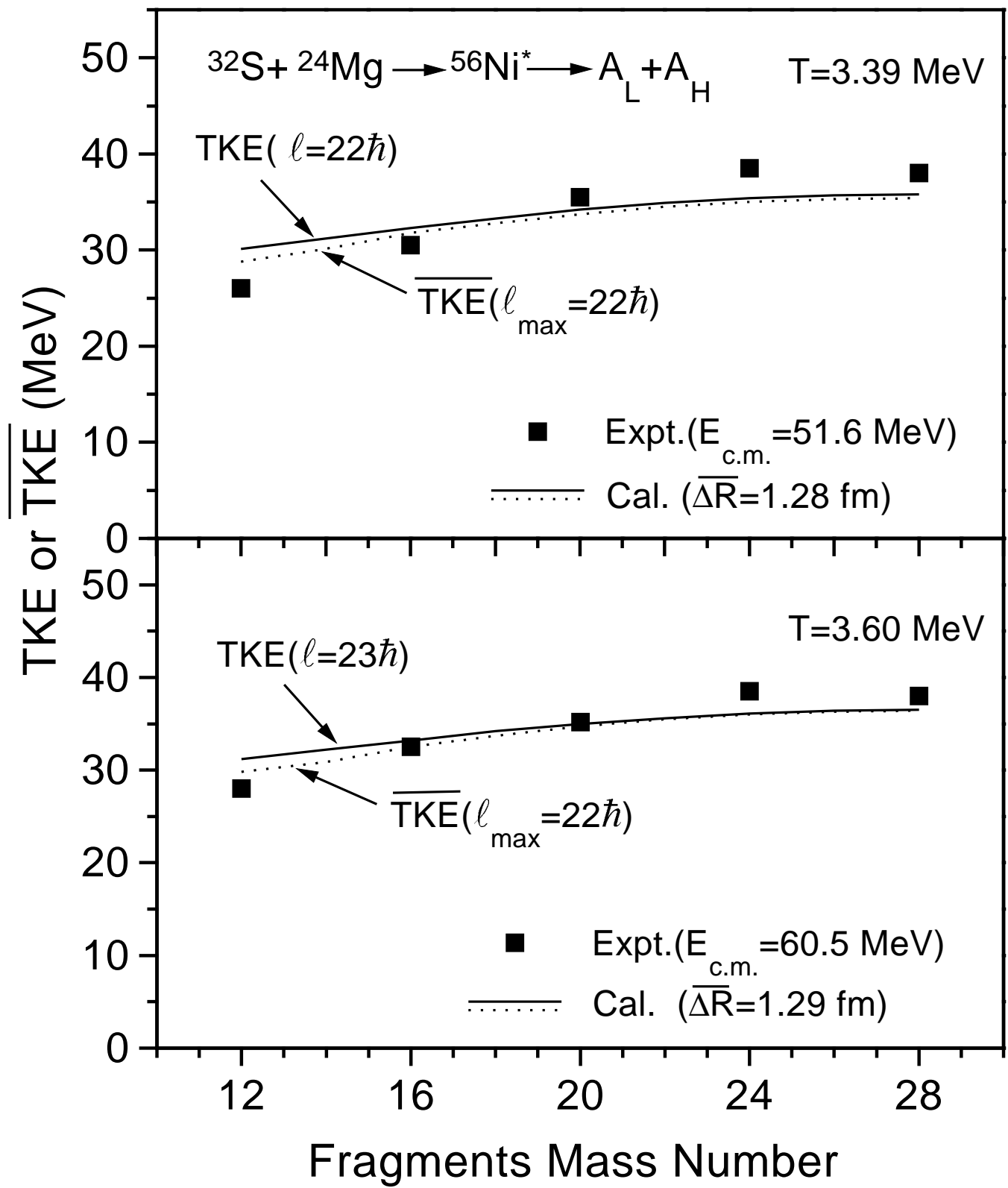
\end{document}